\documentclass[
aps,%
prl,%
reprint,%
amsmath,amssymb,%
floatfix,%
nofootinbib,%
superscriptaddress%
]{revtex4-2}

\usepackage{amsmath,amssymb}
\usepackage{graphicx}
\usepackage{cancel}
\usepackage{comment}
\usepackage[mathscr]{eucal}
\usepackage{epsfig,amsfonts}
\usepackage{bbm}
\usepackage{graphicx}
\usepackage{relsize}
\usepackage{hhline}
\usepackage{diagbox}
\usepackage{psfrag}
\usepackage{mathrsfs} 
\usepackage{array}
\usepackage{tikz}
\usepackage{enumerate}
\usepackage{textcomp}
\usepackage{hyperref}
\usepackage{mathrsfs} 
\usepackage{amsfonts}
\usepackage{lettrine}
\usepackage{multirow}
\usepackage{braket}
\usepackage{microtype}
\usepackage{mathrsfs} 
\usepackage{hyperref}  
\usepackage{xcolor}

\definecolor{myolive}{RGB}{181,204,194}
\definecolor{mypink}{RGB}{222,176,186}
\definecolor{myblue}{RGB}{168,196,215}

\definecolor{myyellow}{RGB}{236,221,171}
\definecolor{myred}{RGB}{210,158,158}
\definecolor{myforestgreen}{RGB}{160,188,176}

\definecolor{mywarmgray}{RGB}{201,194,182}
\definecolor{myterracotta}{RGB}{204,170,150}

\newcommand{\smallcancel}[1]{%
	\mathchoice
	{\ooalign{$\displaystyle#1$\cr\hidewidth$\displaystyle/\hidewidth$\cr}}
	{\ooalign{$\textstyle#1$\cr\hidewidth$\textstyle/\hidewidth$\cr}}
	{\ooalign{$\scriptstyle#1$\cr\hidewidth$\scriptstyle/\hidewidth$\cr}}
	{\ooalign{$\scriptscriptstyle#1$\cr\hidewidth$\scriptscriptstyle/\hidewidth$\cr}}
}
\newcommand{\id}{{\mathbbm{1} }}
\def\be{\begin{equation}}
	\def\ee{\end{equation}}
\def\UniF{a_1}
\def\crp{\rho}

\def\re{{\rm e}}
\def\ri{{\rm i}}

\newcommand{\myDots}{\ifmmode\mathinner{\ldotp\kern-0.2em\ldotp\kern-0.2em\ldotp}\else.\kern-0.13em.\kern-0.13em.\fi}

\def\tr{{\rm Tr}}
\def\SS{L}
\def\bra#1{\langle #1 |}
\def\ket#1{|#1\rangle}

\newcommand{\trten}[4]{
	\draw[thick] (#1+0.25,#2)--(#1+0.25,#2-0.1);
	\draw[thick] (#1+0.25,#2-0.1)to[out=-90,in=-90](#1-0.1,#2-0.1);
	\draw[thick] (#3+0.25,#4+0.5)--(#3+0.25,#4+0.5+0.1);
	\draw[thick] (#3+0.25,#4+0.5+0.1)to[out=90,in=90](#3-0.1,#4+0.5+0.1);
	\draw[thick]  (#3-0.1,#4+0.5+0.1) to (#1-0.1,#2-0.1);
}

\newcommand{\tikzDotsH}[2]{\fill (#1-0.2,#2) circle (1pt)
	(#1,#2) circle (1pt)
	(#1+0.2,#2) circle (1pt);}

\input pdf-trans
\newbox\qbox
\def\usecolor#1{\csname\string\color@#1\endcsname\space}
\newcommand\bordercolor[1]{\colsplit{1}{#1}}
\newcommand\fillcolor[1]{\colsplit{0}{#1}}
\newcommand\outline[1]{\leavevmode%
	\def\maltext{\mydelim #1\mydelim}%
	\setbox\qbox=\hbox{\maltext}%
	\boxgs{Q q 2 Tr \bbthickness\space w \fillcol\space \bordercol\space}{}%
	\copy\qbox%
}
\newcommand\mathbbG[1]{\def\mydelim{$}\outline{#1}}

\newcommand\colsplit[2]{\colorlet{tmpcolor}{#2}\edef\tmp{\usecolor{tmpcolor}}%
	\def\tmpB{}\expandafter\colsplithelp\tmp\relax%
	\ifnum0=#1\relax\edef\fillcol{\tmpB}\else\edef\bordercol{\tmpC}\fi}
\def\colsplithelp#1#2 #3\relax{%
	\edef\tmpB{\tmpB#1#2 }%
	\ifnum `#1>`9\relax\def\tmpC{#3}\else\colsplithelp#3\relax\fi
}
\bordercolor{black}
\fillcolor{white}
\newcommand\bbthickness{.33}

\newcommand{\TenR}[9]{
	\filldraw[fill=myolive,thick, rounded corners=4pt] (#2+0,#3+0) rectangle (#2+#4,#3+#5);
	\draw[thick] (#2,#3+0.5*#5)--(#2-#6,#3+0.5*#5);
	\draw[thick] (#2+#4,#3+0.5*#5)--(#2+#4+#7,#3+0.5*#5);
	\draw[thick] (#2+0.5*#4,#3+#5)--(#2+0.5*#4,#3+#5+#8);
	\draw[thick] (#2+0.5*#4,#3)--(#2+0.5*#4,#3-#9);
	\node[] at (#2+0.5*#4,#3+0.5*#5) {\scalebox{0.8}{$#1$}};}
\newcommand{\NoId}[2]{
	\draw[red,thick] (#1+0.32,#2+0.1)--(#1+0.435,#2+0.26);
}
\newcommand{\TenKR}[9]{
	\filldraw[fill=mypink,thick, rounded corners=4pt] (#2+0,#3+0) rectangle (#2+#4,#3+#5);
	\draw[thick] (#2,#3+0.5*#5)--(#2-#6,#3+0.5*#5);
	\draw[thick] (#2+#4,#3+0.5*#5)--(#2+#4+#7,#3+0.5*#5);
	\draw[thick] (#2+0.5*#4,#3+#5)--(#2+0.5*#4,#3+#5+#8);
	\draw[thick] (#2+0.5*#4,#3)--(#2+0.5*#4,#3-#9);
	\node[] at (#2+0.5*#4,#3+0.5*#5) {\scalebox{0.8}{$#1$}};}

\newcommand{\TenW}[9]{
	\filldraw[fill=myterracotta,thick, rounded corners=4pt] (#2+0,#3+0) rectangle (#2+#4,#3+#5);
	
	\ifdim #6 pt=0pt\relax
	\else
	\draw[thick] 
	(#2,#3+0.5*#5-0.025)--(#2-#6,#3+0.5*#5-0.025);
	
	\draw[thick] 
	(#2,#3+0.5*#5+0.025)--(#2-#6,#3+0.5*#5+0.025);
	
	\draw[thick] 
	(#2-#6*0.5-0.05,#3+0.5*#5+0.025)--
	(#2-#6*0.5,#3+0.5*#5+0.025+0.05);
	
	\draw[thick] 
	(#2-#6*0.5-0.05,#3+0.5*#5-0.025)--
	(#2-#6*0.5,#3+0.5*#5-0.025-0.05);
	\fi

	\draw[thick] (#2+#4,#3+0.5*#5-0.025)--(#2+#4+#7,#3+0.5*#5-0.025);
	\draw[thick] (#2+#4,#3+0.5*#5+0.025)--(#2+#4+#7,#3+0.5*#5+0.025);

	\ifdim #8 pt=0pt\relax
	\else
	\draw[thick,->] 
	(#2+0.5*#4,#3+#5)--(#2+0.5*#4,#3+#5+#8);
	\fi
	\draw[thick] (#2+0.5*#4,#3)--(#2+0.5*#4,#3-#9);

	\node[] at (#2+0.5*#4,#3+0.5*#5) {\scalebox{0.8}{$#1$}};}

\newcommand{\TenKL}[9]{
	\filldraw[fill=myblue,thick, rounded corners=4pt] (#2+0,#3+0) rectangle (#2+#4,#3+#5);
	\draw[thick] (#2,#3+0.5*#5)--(#2-#6,#3+0.5*#5);
	\draw[thick] (#2+#4,#3+0.5*#5)--(#2+#4+#7,#3+0.5*#5);
	\draw[thick] (#2+0.5*#4,#3+#5)--(#2+0.5*#4,#3+#5+#8);
	\draw[thick] (#2+0.5*#4,#3)--(#2+0.5*#4,#3-#9);
	\node[] at (#2+0.5*#4,#3+0.5*#5) {\scalebox{0.8}{$#1$}};}

\begin{document}
	
	\title{Exact strong zero modes are generic in integrable spin systems with large anisotropy}
	
	\author{Sascha Gehrmann}
	\affiliation{The Rudolf Peierls Centre for Theoretical Physics,
		Oxford University, Oxford OX1 3PU, UK}

	\date{\today}
	
	\begin{abstract}
		Strong zero modes (SZMs) are edge-localized operators that commute with the Hamiltonian up to corrections exponentially small in system size, yielding anomalously long edge coherence times. In some settings, notably certain integrable models, this commutator can be made to vanish exactly at finite size, defining an exact SZM (ESZM). Existing ESZM constructions in the integrable setting, however, have proceeded model by model and have not been unified into a common framework. Here, I show that ESZMs arise generically in a broad family of integrable spin models with anisotropic interactions. Their existence follows from two algebraic properties of the underlying R- and K-matrices — quasi-periodicity in the spectral parameter and tracelessness, respectively — providing a uniform, model-independent mechanism. The framework recovers the known ESZM in XXZ chain and its higher-spin generalizations as special cases and predicts ESZMs in previously unrecognized models.
	\end{abstract}

	\maketitle
	
	\paragraph{Introduction.---}
	Strong zero modes (SZMs) are boundary-localized operators that (anti)commute with a discrete global symmetry and commute with the Hamiltonian up to corrections exponentially small in the system size. They enforce a near-degeneracy of every pair of symmetry-related eigenstates and protect edge coherence---features that have made them a sustained focus of attention over the past two decades.
	
	A canonical realization appears in the transverse-field Ising chain, where one can explicitly construct an operator $\bar{\Psi}$ localized near the boundary that anticommutes with the global $\mathbb{Z}_2$ symmetry \cite{kitaev2001unpaired}. Through the Jordan--Wigner transformation, $\bar{\Psi}$ acquires a natural interpretation as a Majorana edge mode. This connection revealed a deep link to topological phases of matter and pointed toward potential applications in robust quantum information processing, thereby catalyzing extensive interest in Majorana zero modes \cite{Alicea_2012}.
	
	Subsequent work generalized the construction of SZMs beyond free-fermion settings to interacting systems and in the presence of disorder \cite{PhysRevB.88.014206,PhysRevResearch.4.L032016,kantha2026strongzeromodesrandom,Kawabata:2017zsb}. Interacting examples include (para)fermionic systems \cite{Bahri15,Alicea:2015hja,sreejith2016parafermion,Moran_2017,Monthus:2018blb,PhysRevB.101.104415,Mahyaeh:2019sxb,Munk:2018oan,vasiloiu2019strong,Svensson:2024djf,Tausendpfund:2025pyy} and cluster or Laplacian models \cite{10.21468/SciPostPhys.14.6.140,katz2025}. 
	They also appear in interfaces between distinct phases \cite{olund2023boundary}, boundary impurities \cite{Yeh:2023cwb} or quantum codes \cite{PRXQuantum.3.020330,Kuno:2023ass}. 
	
	Crucially, these modes are not artifacts of fine-tuning: their dynamical signatures are remarkably robust against perturbations \cite{kemp2017long,PhysRevB.97.064424,yates2019almost} and can even be enhanced by dissipation \cite{PhysRevB.98.094308}. Motivated by the fact that such systems are naturally realized on quantum-simulation platforms, periodically driven (Floquet) settings have become a particularly active venue for SZM research \cite{Sen13,iadecola2015stroboscopic,Khemani16,Yao17,Potter18,Yates:2021asz,mukherjee2024emergent,Vernier24,joshi2026tunablefloquetselectionrules}. This has ultimately led to direct experimental realizations of SZMs \cite{scienceKickedIsing,Jin2025} and opens a path to experimental tests of theoretically predicted effects such as the slowing of entanglement growth \cite{McGinley_2019} and information scrambling \cite{PhysRevB.101.104415}.
	
	Notably, in the integrable XYZ chain, the approximate commutativity with the Hamiltonian can be elevated to an exact one by suitably tuning boundary conditions \cite{Fendley_2016}, yielding exact strong zero modes (ESZMs). More recently, the scope of ESZMs in interacting systems has broadened further. They have been identified in stochastic processes \cite{klobas2023stochastic} and in systems possessing Hilbert-space fragmentation \cite{PhysRevB.101.125126,Datla_2026} and integrable higher-spin systems \cite{Essler2025}. Very recent interesting works report on examples of ESZMs even in non-integrable interacting systems \cite{Moudgalya2026} and at criticality \cite{Prosen:2026knv}. 
	
	Despite this progress, existing constructions remain largely model-specific and rely on case-by-case analyses. Even within integrable settings, it is still unclear which classes of integrable models generically support zero modes and what underlying algebraic structures are responsible for their existence. In this work, I show that exact strong zero modes are in fact a generic feature of integrable spin models with an anisotropy parameter $\Delta$. Their existence can be traced to algebraic properties of the underlying R-matrix and K-matrix, namely that the former is quasi-periodic in the spectral parameter and the latter is traceless. This perspective yields a unified, model-independent framework for understanding ESZMs in integrable spin systems, demonstrating their existence across a broad class of physically interesting models, ranging from previously studied cases such as the XXZ chain
	\begin{align*}
		\mathbbm{H}^{\rm \scalebox{0.5}{XXZ}}=&\sum^{\SS-1}_{j=1}\sigma^x_j\sigma^x_{j+1}+\sigma^y_j\sigma^y_{j+1}+\Delta\sigma^z_j\sigma^z_{j+1}\\
		&+\vec{h}^{\rm \scalebox{0.5}{XXZ}}_{L}\cdot \vec{\sigma}_1+\vec{h}^{\rm \scalebox{0.5}{XXZ}}_{R}\cdot \vec{\sigma}_\SS\,,\notag
	\end{align*}
	to new examples, such as the Izergin–Korepin (IK) chain \cite{Izergin1981}
	\begin{equation}
		\begin{aligned}
			\mathbbm{H}^{\rm \scalebox{0.5}{IK}}=&J\sum^{\SS-1}_{j=0}\sum^8_{\alpha,\beta=1}\lambda_{\alpha,\beta}(\Delta)I^{\alpha}_{j}I^{\beta}_{j+1}\\&+\sum^8_{\alpha=0}h^{\rm \scalebox{0.5}{IK}}_{R,\alpha}I^{\alpha}_{\SS}+h^{\rm \scalebox{0.5}{IK}}_{L,\alpha}\,I^{\alpha}_{1}\,.
			\label{fjdnflsdmsffndjnfdj}
		\end{aligned}
	\end{equation}
	Here, $\sigma^\alpha$ and $I^\alpha$ denote the Pauli and Gell-Mann matrices respectively; $\Delta$ is a free parameter, the so-called anisotropy, and $h$ are external boundary magnetic fields. In the IK model, the couplings $J,\lambda_{\alpha,\beta}(\Delta)$ are fine-tuned functions of $\Delta$ listed in eq.~\eqref{fjdnflsdmsffndjnfdjdsmmm}. While the XXZ chain serves as the paradigmatic model for anisotropic quantum magnetism, the Izergin--Korepin model has found broad application in the statistical mechanics of fluctuating polymer loops and self-avoiding walks, through its correspondence with dilute $O(n)$ loop models \cite{Nienhuis1982,Nienhuis1990,Yung1995,Vernier2014,Vernier2015}.
	
	The manuscript begins with a brief review of the general framework in which a wide class of integrable spin chains, including the models above, can be formulated in a unified manner: the transfer-matrix formalism and the Quantum Inverse Scattering Method (QISM) \cite{FaddeevTakhtajan1979,KorepinBogoliubovIzergin1993}. Then, after introducing a precise definition of ESZMs, the manuscript proceeds to present their general construction in this framework and ends with a brief numerical demonstration of the infinite edge coherence time in the IK chain \eqref{fjdnflsdmsffndjnfdj} due to the presence of ESZMs.
	
	\paragraph{Integrable systems.---}
	Consider an one-dimensional quantum system with Hilbert space  $\mathscr{H} = \mathscr{V}^{\otimes \SS}$
	where $\SS$ denotes the system size and $\mathscr{V}$  
	is a local Hilbert space of dimension $N = \mathrm{dim}(\mathscr{V})$.
	The two fundamental objects for an integrable system with open boundary conditions are the R-matrix $R(u)$ and the K-matrix $K^-(u)$, endomorphisms on $\mathscr{V}\otimes \mathscr{V}$ and $\mathscr{V}$ respectively
	depending on the so-called spectral parameter $u\in \mathbb{C}$. 
	Denoting by $O_{i,j}$ the operators acting
	non-trivially on the $i$-th and $j$-th 
	tensorial factors of the whole Hilbert space $\mathscr{H}$ and as the identity on the 
	remaining spaces, the defining equation of $R(u)$ and $K^-(u)$, the Yang-Baxter equation (YBE) \cite{Yang1968,Baxter1972,Baxter1982} and Sklyanin's reflection equation (SRE) \cite{Cherednik1984,Sklyanin_1988}, read
	\begin{align}
		R_{1,2}(u-v)&R_{1,3}(u)R_{2,3}(v)\notag\\
		&=R_{2,3}(v)R_{1,3}(u)R_{1,2}(u-v)\,,\\
		R_{1,2}(u-v)&K^{-}_{1}(u)R_{2,1}(u+v)K^{-}_2(v)\notag\\
		&= K^{-}_2(v)R_{1,2}(u+v)K^{-}_{1}(u)R_{2,1}(u-v)\,.\label{mfkfdkmfdkmf}
	\end{align}
	For the reader's convenience, the explicit R- and K-matrices studied here are listed in the end matter, which correspond to the R-matrices found by Jimbo and Bazhanov \cite{Jimbo1986, Bazhanov1987} and the K-matrices found by Lima-Santos \cite{Lima_Santos_All}.  They depend on a free parameter $q$, also often rewritten as the anisotropy $\Delta=\frac{1}{2}(q+q^{-1})$. 
	Given a solution of the YBE and SRE, one can define a transfer matrix 
	\begin{equation}
		\begin{aligned}
			\mathbb{T}(u)&=\tr_0(K^+_0(u)T_0(u)K^-_0(u)\hat{T}_0(u))\,,\\
			T_0(u)&=R_{0,1}(u)\dots R_{0,\SS}(u),\\
			\hat{T}_0(u)&=R_{\SS,0}(u)\dots R_{1,0}(u),
		\end{aligned}
	\end{equation}
	where $K^{+}(u)=\bigl(K^{-}(-u-\crp)\bigr)^{t} M$, with $M$ and $ \crp $ denoting model-dependent symmetry data associated with the R-matrix, whose explicit expressions are also collected in the end matter \cite{note1}. 
	Due to the YBE and SRE the transfer matrix commutes with itself for different arguments $\left[\mathbb{T}(u),\mathbb{T}(v)\right]=0$.
	By expanding the transfer matrix at $u=0$, an extensive number of mutually commuting operators local $\mathbb{Q}^{(m)}$. Picking any - most commonly $\mathbb{Q}^{(1)}$ - as the Hamiltonian, one obtains a system with an extensive number of conserved charges, a hallmark of integrability \cite{Caux_2011}. One should stress that these conserved charges obtained in this way are sums of local densities of interaction range $m+1$ in the bulk and $m$-site boundary interaction. In the following, this form is referred as ultra local. For the concrete examples of the XXZ or IK- Hamiltonians in \eqref{fjdnflsdmsffndjnfdj} the R-matrix is given as the $A^{(1)}_{1}$ and the $A^{(2)}_{2}$ ones in the end matter respectively.
	
	To ease notation, it is useful to represent the R- and K-matrices graphically
	\begin{equation}
		\begin{tikzpicture}[baseline={(0,0.68)}]
			\node [right] at (-1.1,0.75) {$R(u)=$};
			\draw[thick,->] (0.75,0.25)--(0.75,1.25);
			\draw[thick,->] (0.5,0.75)--(1.25,0.75);
			
			\TenR{u}{0.5}{0.5}{0.5}{0.5}{0.25}{0.25}{0.25}{0.25}
			
			\node [right] at (-1.1+2.75,0.75) {$K^+(u)=$};
			\draw[thick,<-] (0.5+3,0.25)--(0.5+3,1.25);
			\TenKL{u}{3.25}{0.5}{0.5}{0.5}{0}{0}{0.25}{0.25}
			\node [right] at (-1.1+5.25,0.75) {$K^-(u)=$};
			\draw[thick,->] (6,0.25)--(6,1.25);
			\TenKR{u}{5.75}{0.5}{0.5}{0.5}{0}{0}{0.25}{0.25}
		\end{tikzpicture}\,.
	\end{equation}
	Then the transfer matrix can be represented as 
	\begin{equation}\label{jfdjnfjdnfjfn}
		\begin{tikzpicture}[baseline={(0,0.68)}]
			\draw[thick,dashed] (1,0.25)--(6,0.25);
			\draw[thick,dashed] (1,0.25+1.)--(6,0.25+1.);
			\draw[thick,>->] (6,0.315)--(6,1.175);
			\draw[thick,<-<] (0.75,0.315)--(0.75,1.175);
			\draw[thick,->] (1.25,0.5)--(1.25,1.76);
			\draw[thick,->] (2.,0.5)--(2.,1.76);
			\draw[thick,->] (2.75,0.5)--(2.75,1.76);
			\draw[thick,->] (5.5,0.5)--(5.5,1.76);

			\node [right] at (-1.1,0.75) {$\mathbb{T}(u)=$};
			
			\TenKL{u}{0.5}{0.5}{0.5}{0.5}{0}{0}{0.2625}{0.2625}
			
			\TenR{u}{1.00}{0}{0.5}{0.5}{0.25}{0.25}{0.25}{0.25}
			\TenR{u}{1.00}{1}{0.5}{0.5}{0.25}{0.25}{0.25}{0.25}
			\TenR{u}{1.75}{0}{0.5}{0.5}{0.25}{0.25}{0.25}{0.25}
			\TenR{u}{1.75}{1}{0.5}{0.5}{0.25}{0.25}{0.25}{0.25}
			\TenR{u}{2.50}{0}{0.5}{0.5}{0.25}{0.25}{0.25}{0.25}
			\TenR{u}{2.50}{1}{0.5}{0.5}{0.25}{0.25}{0.25}{0.25}
			
			\tikzDotsH{4.0}{0.75}
			
			\TenR{u}{5.25}{0}{0.5}{0.5}{0.25}{0.25}{0.25}{0.25}
			\TenR{u}{5.25}{1}{0.5}{0.5}{0.25}{0.25}{0.25}{0.25}
			\TenKR{u}{5.75}{0.5}{0.5}{0.5}{0}{0}{0.2625}{0.2625}
		\end{tikzpicture}\,.
	\end{equation}
	
	\paragraph{Exact strong zero modes.---}
	In this manuscript, an ESZM in the setting of the QISM is defined as an operator  
	$\mathbbG{\Psi}$ satisfying the following three properties.  
	
	\textit{(i) Exact commutativity.}  
	The operator commutes with the generating functional of conserved quantities of the integrable model,  
	namely the transfer matrix $\mathbbm{T}(u)$,
	\begin{align}
		[\mathbbm{T}(u), \mathbbG{\Psi}] = 0\,.
	\end{align}
	
	\textit{(ii) Localization.}  
	$\mathbbG{\Psi}$ can be expressed as
	\begin{align}\label{fösanfjnfjnjrrffr}
		\mathbbG{\Psi} = \sum_{j=1}^{\SS} \Psi_j\,,
	\end{align}
	where $\Psi_j$ acts as the identity on all sites $k > j$, non-trivially on site $j$, and arbitrarily on sites $k < j$. By definition, $\Psi_j$ parametrizes the component of the operator with non-trivial support restricted to sites $\leq j$.
	One requires the exponential decay of the Hilbert--Schmidt norms
	\begin{align}
		\|\Psi_j\|^2 = N^{-\SS} \, \mathrm{tr}(\Psi_j^\dagger \Psi_j)\,
	\end{align}
	in the thermodynamic limit $\SS \to \infty$, namely,
	\begin{align}
		\lim_{\SS\to \infty}\|\Psi_j\|^2 \propto \re^{-\alpha j}\,,
	\end{align}
	for some $\alpha > 0$.  This means that the operators' support is mainly concentrated at the boundary, i.e., it is localized at the boundary.
	
	\textit{(iii) Normalization and global action.}  
	The operator is normalizable in the thermodynamic limit, meaning that $\lim_{L\to \infty}\|\mathbbG{\Psi}\|^2 = 1$ exists. Further, one imposes that $\mathbbG{\Psi}$ acts non-trivially across the full Hilbert space,  
	rather than within a restricted energy sector.  
	A possible condition is to require $
	\mathbbG{\Psi}^n \propto \id $ as $ \SS \to \infty$
	for some integer $n$. However, the following construction is inherently not specific to any energy regime.

	\paragraph{The general construction of ESZMs.---}
	The expansion of the transfer matrix at $u=0$ yields the ultra local conversed charges. To obtain other conserved operators with different locality properties like the ESZM, it is natural to consider a different expansion point.
	
	This has been done for the XXZ model \cite{Vernier24,Fendley2025,Gehrmann2025}, and for its higher-spin generalizations \cite{Essler2025}, where multiple ESZMs were shown to coexist \cite{note4}.  Here I extend these results, showing that ESZMs exists in the broad class of integrable models classified by the affine Lie algebras $A^{(1)}_{n-1}$, $A^{(2)}_{2n-1}$, $A^{(2)}_{2n}$, $B^{(1)}_{n}$, $C^{(1)}_{n}$, and $D^{(1)}_{n}$ \cite{Jimbo1986, Bazhanov1987}.
	
	The underlying mechanism for the existence of ESZM in such a broad class of models can be related to the quasi-periodicity of the R-matrix. Many R-matrices are known to be (quasi)-periodic with period $p\in\mathbb{C}$, meaning that   
	\begin{align}\label{pamfbelwmsjfnfhsh}
		R_{1,2}(u+p)=\pm U_1 R_{1,2}(u)U^{-1}_1
	\end{align}
	whereby $U$ is an invertible matrix and $U=\id$ for pure periodicity.
	In the following, I restrict the analysis to R-matrices found by Jimbo, yielding pure periodicity with period $p=2\ri \pi$. They are explicitly presented in the end matter. Note that the pure periodicity can be easily seen as all appearing functions $a_i(u),b^\pm(u),c^\pm(u), d^\pm(u)$ are $2\ri \pi$-periodic, see Eqs.~\eqref{rjnjsndjsnjdsn},\eqref{tjenjsnfjyll}-\eqref{kgdfkdnfjdnfjdnjfdn}. One further finds $R_{12}(\frac{p}{2})\neq \id,P_{12}$, where $P$ is the permutation operator. If the R-matrix is in a different gauge, the same following formalism applies with slight modifications \cite{note2}.
	
	Consider now the impact of the periodicity of the R-matrix on the K-matrix.  Let $K^-(u\,|\,\{\beta^-\})$ be a solution of the SRE, possibly featuring some free parameters $\{\beta^-\}$. By performing the shift
	$u \mapsto u+\tfrac{p}{2}, v \mapsto v+\tfrac{p}{2}$
	in the SRE and exploiting the periodicity of the R-matrix, it follows immediately that
	$\tilde{K}^-(u\,|\, \{\beta^-\}) := K^-\left(u+\tfrac{p}{2}\,|\, \{\beta^-\}\right)$
	is again a solution of the reflection equation. At this point, two possibilities arise. Either $\tilde{K}^-$ defines a \emph{new} solution of the reflection equation, or there exists a mapping $f_u \,:\,  (\{\beta^-\})\mapsto (\{\tilde{\beta}^-\})$  of its free parameters $\{\beta^-\}$ onto another set $\{\tilde{\beta}^-\}$ such that it is proportional to the original one
	$\tilde{K}^{-}(u\,|\,\{\beta^-\}) \propto K^{-}(u\,|\, \{\tilde{\beta}^-\})$ modulo symmetry operations. In this case, one simply recovers the well-known scale symmetry of the reflection equation,
	$K^-(u) \mapsto g(u)K^-(u)$,
	with $g(u)$ an arbitrary scalar function.
	
	The first possibility would be quite strong: it would imply that new solutions could be generated solely by exploiting the periodicity of the R-matrix. Indeed, the empirical evidence is quite convincing:  by a direct inspection of the K-matrices classified in \cite{Lima_Santos_A1n,Lima_Santos_B1n_A22n,Lima_Santos_C1n_D1n_A22n1,Lima_Santos_D2n}
	and summarized in \cite{Lima_Santos_All}, which exhausts a vast majority of all known K-matrices, one finds that the shifted solution $\tilde K$ is indeed proportional to the original one by a reparametrisation of the free parameters and additional symmetry transformations. This implies that if one assumes the regularity of the K-matrix at $u=0$ meaning,
	\begin{align}
		K^-(0\,|\,\{\beta^-\})\propto \id\,,
	\end{align}
	and so all dependence on free parameters drops out, one obtains \cite{note2}
	\begin{align}\label{fdjfdjfpaosos}
		K^-(\tfrac{p}{2}\,|\,\{\beta^-\})\propto K^-(0\,|\,\{f_0(\beta^-)\})\propto\mathbbm{1}\,.
	\end{align}
	
	Now let us discuss how this property allows one to obtain an operator with the spatial expansion of the kind \eqref{fösanfjnfjnjrrffr}. For this, one also needs the so-called unitarity property of the R-matrix  $R_{1,2}(v)R_{2,1}(-v)= \UniF(v)\UniF(-v)\id$ where $\UniF(v)$ is a scalar function. Combining these two properties yields the following pull-through identity: 
	\begin{equation}\label{djsdnjsndjlsiurnfbsdnsn}
		\begin{aligned}
			R_{0,j}(\tfrac{p}{2}+v)&K^{-}_0(\tfrac{p}{2})R_{j,0}(\tfrac{p}{2}-v)\\&\propto R_{0,j}(-\tfrac{p}{2}+v)R_{j,0}(\tfrac{p}{2}-v)\propto\id
		\end{aligned}
	\end{equation}
	or graphically
	\begin{equation}
		\begin{tikzpicture}[baseline={(0,0.68)}]
			\draw[thick, ->] (0.125,0)--(0.125,1.75);
			\draw[thick, <-] (-0.5,1.25)--(0,1.25);
			\TenR{\tfrac{p}{2}\pm v}{-0.25}{0}{0.75}{0.5}{0.25}{0.25}{0.25}{0.25}
			\TenR{\tfrac{p}{2}\mp v}{-0.25}{1}{0.75}{0.5}{0.25}{0.25}{0.25}{0.25}
			\TenKR{\tfrac{p}{2}}{0.5}{0.5}{0.5}{0.5}{0}{0}{0.2625}{0.2625}
			\node[] at (1.5,0.75) {$\propto$};
			
			\def\x{2.25}
			\draw[thick, ->] (0.5+\x,0)--(0.5+\x,1.75);
			\draw[thick, <-] (-0.25+\x,1.25)--(0+\x,1.25);
			\TenR{\tfrac{p}{2}\pm v}{0+\x}{0}{1.0}{0.5}{0.25}{0.25}{0.25}{0.25}
			\TenR{-\tfrac{p}{2}\mp v}{0+\x}{1}{1}{0.5}{0.25}{0.25}{0.25}{0.25}
			\draw[thick] (1.24+\x,0.25)--(1.24+\x,1.25);
			\node[] at (1.75+\x,0.75) {$\propto$};
			
			\def\x{2.0}
			\draw[thick, ->] (0.425+2.75+\x,0)--(0.425+2.75+\x,1.75);
			\draw[thick, <-] (-0.25+2.75+\x,1.25)--(0+2.75+\x,1.25);
			\draw[thick, -] (-0.25+2.75+\x,1.25-1)--(0+2.75+\x,1.25-1);
			\draw[thick] (0.75+\x+2,0.25)--(0.75+\x+2,1.25);
		\end{tikzpicture}\,\,,
	\end{equation}
	where \eqref{pamfbelwmsjfnfhsh} and \eqref{fdjfdjfpaosos} were applied in the first and unitarity of the R-matrix in the last step, and $v$ is arbitrary. 
	
	The pull-through relation implies then that the first derivative of the transfer matrix $\mathbbm{T}'(\tfrac{p}{2})$ at $u=\tfrac{p}{2}$ has the form 
	\begin{align}\label{fjdapedmdjensjsbrjgbfb}
		\mathbbm{T}'(\tfrac{p}{2})=\sum^{L+1}_{j=1}\tilde{\Psi}_j\,.
	\end{align}
	The $\tilde{\Psi}_k$ terms originate from the derivative acting on the R-matrices at site $k$, such that the pulling-through identity terminates there  (and $k=L+1$ is the term where the derivative hits the right K-matrix). Further, note that as $R_{12}(\frac{p}{2})\neq \id,P_{12}$, ultra-locality does not hold. Consequently, $\tilde{\Psi}_{k}$ is not confined to a finite neighbourhood of the site $k$, but generally features a support extending over the entire interval from $1$ to $k$. The above facts already provide a strong indication that $\mathbb{T}'\!\left(\tfrac{p}{2}\right)$ may be localized at the boundary. 
	
	The above finding motivates defining an ESZM $\mathbbG{\Psi}$ for integrable spin models as follows:
	\begin{align}\label{fnjnjnjnjnjfnff}
		\mathbbG{\Psi}=\mathcal{N}\, \left(\mathbb{T}'\big(\tfrac{p}{2}\big)-\frac{\tr_{\mathcal{H}}\left(\mathbb{T}'\big(\tfrac{p}{2}\big)\right)}{{\rm dim}(\mathcal{H})}\,\id\right)
	\end{align}
	where $\mathcal{N}$ is a normalisation constant ensuring that the Hilbert-Schmidt (HS) norm is one
	\begin{align}
		|\!|\mathbbG{\Psi}|\!|^2=N^{-\SS}\,\tr(\mathbbG{\Psi}^\dagger\mathbbG{\Psi})=1\,.
	\end{align}
	The additive constant is chosen so that $\mathbbG{\Psi}$ has zero overlap with the identity operator on the whole Hilbert space. As the ESZM is defined as the derivative of the transfer matrix $\mathbbm{T}(u)$, it clearly commutes exactly with $\mathbbm{T}(u)$. Further, there was no restriction to any energy regime, implying global action. Now let's turn to the final property, spatial locality. 
	\paragraph{Localisation at the boundary.---}
	\begin{figure}[t]
		\centering
		\includegraphics[width=\linewidth]{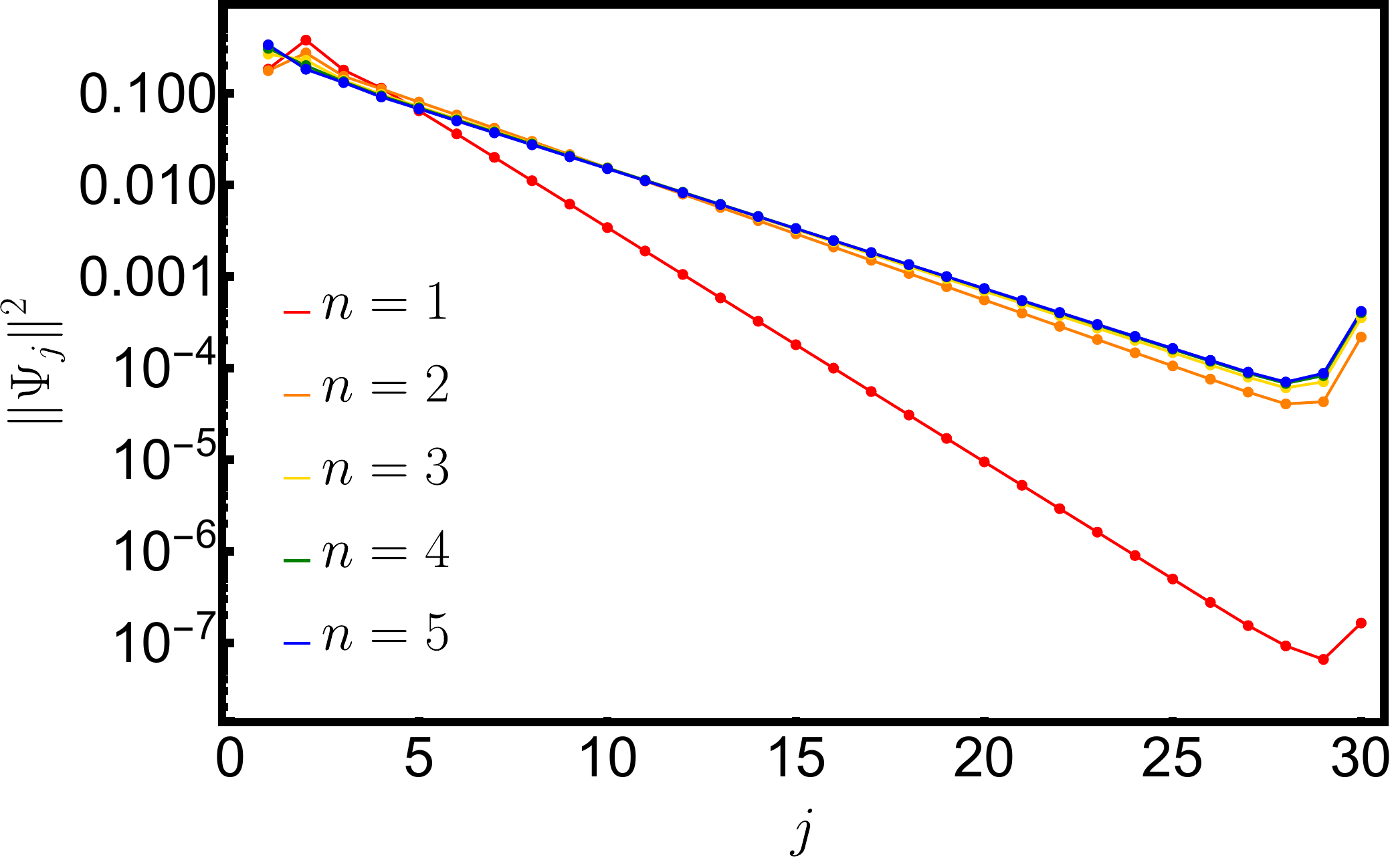}
		\caption{
			Exponential decay of the Hilbert-Schmidt norms of the contributions $\Psi_j$ to the ESZM for the $A^{(2)}_{2n}$ series for $n=1,2,3,4,5$. Here the R-matrix is given by \eqref{kfefmdnfdjflafndfnf}, $\Delta=1.811$, and the $K^-$ is chosen as \eqref{ksmdgnfjernenjd2n} and $K^+$ is fixed by isomorphism. The parameters are taken as  $\beta^+_{1,1}=\frac{1}{2}$, $\beta^+_{i,i'}=0.1$, $\beta^-_{1,1}=-0.2$, $\beta^-_{i,i'}=1.1$. Interestingly, one sees that the lower rank case $n=1$ decays way faster than the other higher rank models.
			\label{fig:An1}}
	\end{figure}
	Having established a general pulling-through relation, it is clear that $\mathbb{T}'(\tfrac{p}{2})$ takes the form~\eqref{fjdapedmdjensjsbrjgbfb}, and likewise $\mathbbG{\Psi}$ in~\eqref{fnjnjnjnjnjfnff}. However, it remains to quantify the locality of the terms appearing in the sum, as these need not exhibit decay.
	
	It is important to notice that the quantities $\tilde{\Psi}_j$ may still contain contributions that act as the identity on site $j$. It is therefore necessary to project out any remaining operators contained in $\tilde{\Psi}_j$ that act trivially on site $j$ by defining
	\begin{align}
		\Psi_j=\frac{1}{N^{L-j}}\tr_{j+1,\dots N} \left(\mathbbG{\Psi}\right)-\frac{1}{N^{L-j+1}}\tr_{j,\dots \SS} \left(\mathbbG{\Psi}\right)\,.
	\end{align}
	By standard techniques within the Matrix Product Operator (MPO) formalism, one can efficiently evaluate the HS norms $|\!|\Psi_j|\!|^2$ for a given R-matrix for intermediate system sizes. In Fig.~\ref{fig:An1}, I present results for the first few representatives of the $A^{(2)}_{2n}$ series, which clearly exhibit exponential decay. Similar behavior is observed across all studied series  \cite{Supplemental}.
	Based on a numerical study, I find consistent exponential decay provided the two sufficient conditions are obeyed. First, the anisotropy parameter $\Delta$ needs to obey $|\Delta|>1$ (for $B^{(1)}_n$ a slightly stronger condition needs to apply, see \cite{Supplemental}).  Secondly, the boundary conditions at which the ESZM (here, the left one) is expected to be localized must be fixed in the following manner. The dual vector \begin{align}
		\bra{\mathcal{K}}
		=&\sum^N_{\{k_i\}=1}[(K^+)^*(\tfrac{p}{2})]^{k_2}_{k_1} [K^+(\tfrac{p}{2})]^{k_4}_{k_3}\bra{k_1,k_2,k_3,k_4}
	\end{align}
	must be orthogonal 
	\begin{align}\label{kdskidjndeufjfndjhbfdhjn}
		\braket{\mathcal{K}|\mathcal{R}_0}=0
	\end{align}
	to the tensor product of the two maximal entangled Bell states 
	\begin{align}\label{jdnjdnfjdnfjdhskak}
		\ket{\mathcal{R}_0}=\frac{1}{N} \left(\sum^{N}_{j=1} \, \ket{j,j}\, \right)\otimes \left(\sum^{N}_{j=1} \, \ket{j,j}\, \right)\,.
	\end{align}
	The latter is the right eigenstate associated with the largest eigenvalue of the transfer operator
	\begin{align}
		\mathcal{W}^{\mathbf{p}}_{\mathbf{k}}=&\sum_{\alpha,\beta,\gamma,\delta}
		[R^*(\tfrac{p}{2})]^{p_1\gamma  }_{k_1\beta}[R^*(\tfrac{p}{2})]^{\beta k_2}_{\alpha p_2}
		[R(\tfrac{p}{2})]^{p_3\gamma}_{k_3\delta}[R(\tfrac{p}{2})]^{\delta k_4}_{\alpha p_4}\notag
	\end{align}
	appearing in the calculation of the HS norms of $\Psi_j$\cite{Supplemental}. I point out that I find empirically that $\ket{\mathcal{R}_0}$ is the same for all considered integrable models, only dependent on the local Hilbert space dimension $N$. Note that the orthogonality condition reduces simply to 
	\begin{align}\label{fkjdnfjdnjfndjfskk}
		\tr\Big(K^+\big(\tfrac{p}{2}\big)\Big)=0\,.
	\end{align}
	Suitable BCs obeying \eqref{kdskidjndeufjfndjhbfdhjn}/\eqref{fkjdnfjdnjfndjfskk} for the considered cases are listed in the end matter. 
	
\begin{figure}[t]
		\centering
		\includegraphics[width=\linewidth]{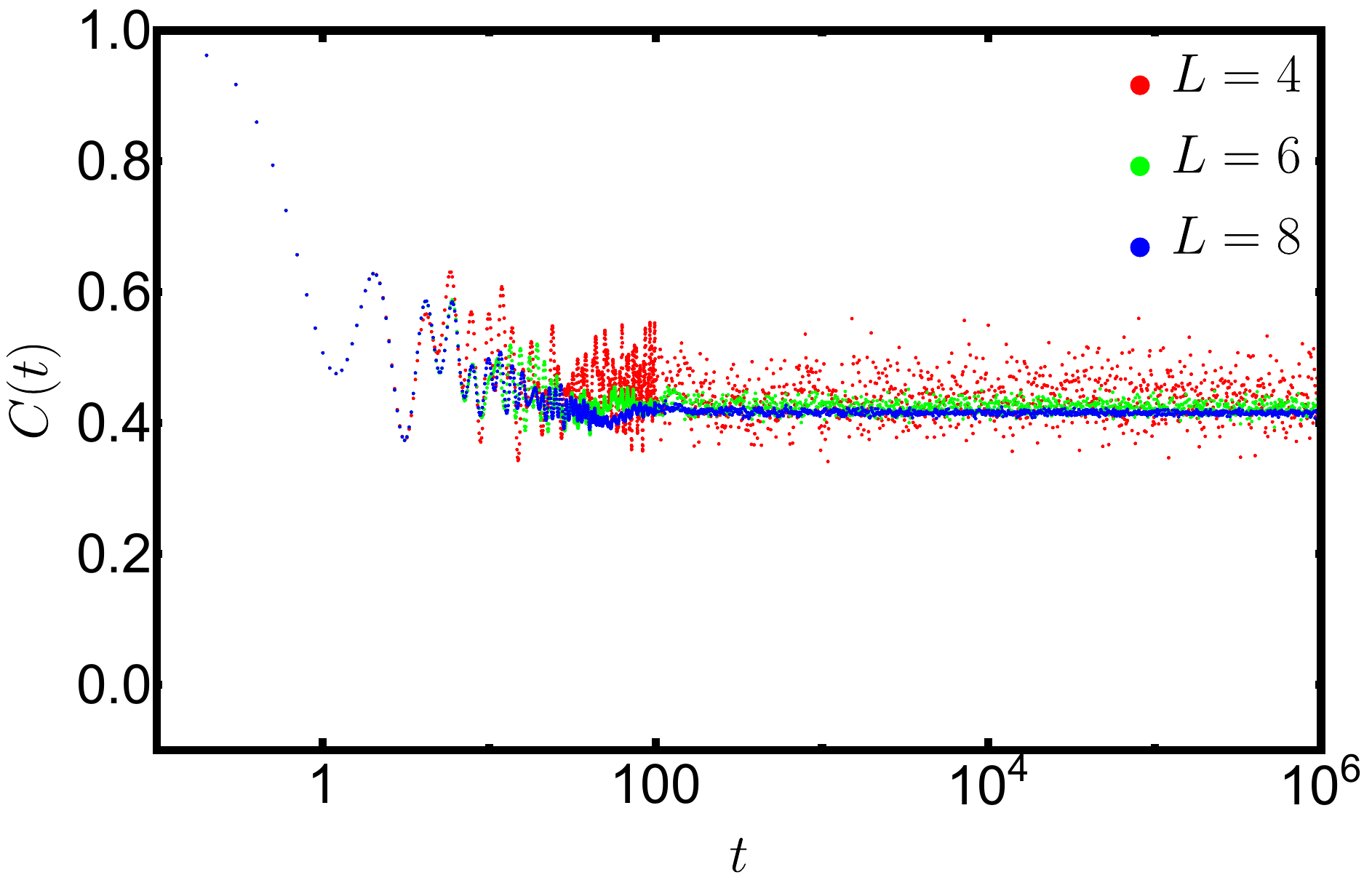}
		\caption{Infinite-temperature autocorrelation function $C(t)$ of the boundary observable $I^4_1 - S^z_1$---a spin-$\tfrac{1}{2}$ in a field tilted $45^\circ$ from the $z$-axis, restricted to the extremal-magnetization sector $\pm 1$ of site $1$---evolved under the IK chain Hamiltonian~\eqref{fjdnflsdmsffndjnfdj}, computed by exact diagonalization on small system sizes. The autocorrelator saturates at a finite, non-vanishing long-time value, signalling the presence of ESZMs. Parameters: $\Delta = 2.5$; the non-vanishing components of the left-boundary magnetic field are $h^{\rm\scalebox{0.5}{IK}}_{R,3} = 0.700$, $h^{\rm\scalebox{0.5}{IK}}_{R,4} = -3.061$, $h^{\rm\scalebox{0.5}{IK}}_{R,8} = -3.464$, while the right-boundary field vanishes.
			\label{fig:AutoCorr}}
	\end{figure}

	\paragraph{Application: edge protection in the IK-chain.---}
	ESZM are expected to give rise to an infinite edge coherence time: the autocorrelation function of a boundary-localized observable, under evolution by any member of the commuting family, is protected by the ESZMs and saturates at a long-time value set by its overlap with all ESZMs in the system \cite{Gehrmann2025}. This is demonstrated in Fig.~\ref{fig:AutoCorr} for the autocorrelation function in the infinite temperature state
	\begin{align}
		C(t)=N^{-L}\,\tr\big[\mathcal{O}(t)\mathcal{O}(0)\big]
	\end{align}
	under the time evolution by IK Hamiltonian~\eqref{fjdnflsdmsffndjnfdj},
	obtained by exact diagonalization on small system sizes. The observable is taken to be $\mathcal{O}=I^4_1 - S^z_1$, which acts nontrivially only on the $\{|+1\rangle, |-1\rangle\}$ subspace of the boundary site, where it reduces to $\sigma^x - \sigma^z$---a spin-$\tfrac{1}{2}$ in a field tilted $45^\circ$ from the $z$-axis on the extremal-magnetization sector.

	\paragraph{Conclusion.---}
	I presented a generic mechanism for constructing exact strong zero modes (ESZMs) in a broad class of integrable models, namely, spin models associated with the solutions of the Yang–Baxter equation for the Lie algebra series 
	$A^{(1)}_{n-1}$, $A^{(2)}_{2n-1}$, $A^{(2)}_{2n}$, $B^{(1)}_{n}$, $C^{(1)}_{n}$ and $D^{(1)}_{n}$. 
	The key observation underlying the construction is that any integrable model in this class has a distinguished expansion point $u=\frac{p}{2}$ identified from the (quasi-)periodicity of the R-matrix. Combined with unitarity, this property leads to a general pull-through relation that 
	produces the spatial structure required for boundary-localization. Empirically, I identified an explicit sufficient criterion \eqref{kdskidjndeufjfndjhbfdhjn} on the allowed boundary conditions under which the ESZMs are localized at the edge. The resulting localization properties were demonstrated numerically for the first few representatives of the considered Lie algebra series. 
	
	The goal of this work was not to construct all ESZMs for each individual model, but rather to demonstrate that their existence is a generic feature of integrable systems with  $|\Delta|>1$ (and a slightly stronger condition for $B^{(1)}_n$). A systematic and exhaustive construction of all ESZMs remains an open problem for future work. It would be interesting to understand the present construction in alternative approaches, such as commutant algebras \cite{
		PhysRevB.107.224312,PRXQuantum.5.040330,MOUDGALYA2023169384,Moudgalya2026} or telescoping and intertwining relations \cite{frassek2026intertwiningmarkovprocessesmatrix,rubio2026localcharacterizationglobaltensor,Fendley2025,Gehrmann:2026ypf}. Since strong zero modes have also been constructed for the XYZ chain \cite{Fendley_2016}, and later systematized and generalized to arbitrary boundary conditions in \cite{Fendley2025}, it is natural to expect that the mechanism proposed here extends to the higher-rank elliptic case. Further, as $v$ is arbitrary in \eqref{djsdnjsndjlsiurnfbsdnsn}, one can also study ESZM in the integrable quantum circuits associated with the R-matrices ~\cite{Vernier24,Gehrmann2025}.
	
	Two other natural extensions are the physical consequences of infinite edge coherence in the IK model, via its connection to loop and $O(n)$ models, and the Hamiltonian limit of the remaining models, where ESZMs are expected to produce infinite edge coherence times.
	
	\paragraph{Acknowledgments.---}  {This work was supported by the EPSRC under grant EP/X030881/1. I thank Dumitru C\u{a}lug\u{a}ru, Fabian Essler, Paul Fendley, Lorenzo Piroli, Ana L. Retore and Rafael I. Nepomechie for useful discussions.  }
	
	\bibliographystyle{apsrev4-2}
	\bibliography{references}

@article{kemp2017long,
  title={Long coherence times for edge spins},
  author={Kemp, Jack and Yao, Norman Y and Laumann, Christopher R and Fendley, Paul},
  journal={Journal of Statistical Mechanics: Theory and Experiment},
  volume={2017},
  number={6},
  pages={063105},
  year={2017},
  publisher={IOP Publishing},
  doi={/10.1088/1742-5468/aa73f0}
}

@article{olund2023boundary,
  title = {Boundary strong zero modes},
  author = {Olund, Christopher T. and Yao, Norman Y. and Kemp, Jack},
  journal = {Phys. Rev. B},
  volume = {111},
  issue = {20},
  pages = {L201114},
  numpages = {6},
  year = {2025},
  month = {May},
  publisher = {American Physical Society},
  doi = {10.1103/PhysRevB.111.L201114}
}

@article{Jin2025,
  author = {Jin, Feitong and Jiang, Si and Zhu, Xuhao and Bao, Zehang and Shen, Fanhao and Wang, Ke and Zhu, Zitian and Xu, Shibo and Song, Zixuan and Chen, Jiachen and et al},
  title = {Topological prethermal strong zero modes on superconducting processors},
  journal = {Nature},
  year = {2025},
  volume = {645},
  number = {8081},
  pages = {626--632},
  doi = {10.1038/s41586-025-09476-z},
  url = {https://doi.org/10.1038/s41586-025-09476-z},
  issn = {1476-4687}
}

@article{kitaev2001unpaired,
  title={Unpaired Majorana fermions in quantum wires},
  author={Kitaev, A Yu},
  journal={Physics-uspekhi},
  volume={44},
  number={10S},
  pages={131},
  year={2001},
  doi={10.1070/1063-7869/44/10S/S29},
  publisher={IOP Publishing}
}

@article{vasiloiu2019strong,
    author = "Vasiloiu, Loredana M. and Carollo, Federico and Marcuzzi, Matteo and Garrahan, Juan P.",
    title = "{Strong zero modes in a class of generalized Ising spin ladders with plaquette interactions}",
    primaryClass = "cond-mat.stat-mech",
    doi = "10.1103/PhysRevB.100.024309",
    journal = "Phys. Rev. B",
    volume = "100",
    number = "2",
    pages = "024309",
    year = "2019"
}

@article{klobas2023stochastic,
    author = "Klobas, Katja and Fendley, Paul and Garrahan, Juan P.",
    title = "{Stochastic strong zero modes and their dynamical manifestations}",
    primaryClass = "cond-mat.stat-mech",
    doi = "10.1103/PhysRevE.107.L042104",
    journal = "Phys. Rev. E",
    volume = "107",
    number = "4",
    pages = "L042104",
    year = "2023"
}

@article{Reshetikhin1990,
  author  = {Reshetikhin, N. Yu. and Semenov-Tian-Shansky, M. A.},
  title   = {Central extensions of quantum current groups},
  journal = {Letters in Mathematical Physics},
  year    = {1990},
  volume  = {19},
  number  = {2},
  pages   = {133--142},
  doi     = {10.1007/BF01045884},
  url     = {https://doi.org/10.1007/BF01045884},
  issn    = {1573-0530}
}

@article{Bazhanov1987,
  author       = {Bazhanov, Vladimir V.},
  title        = {Integrable quantum systems and classical Lie algebras},
  journal      = {Communications in Mathematical Physics},
  year         = {1987},
  volume       = {113},
  number       = {3},
  pages        = {471--503},
  issn         = {1432-0916},
  doi          = {10.1007/BF01221256},
  url          = {https://doi.org/10.1007/BF01221256},
}

@article{Jimbo1986,
  author    = {Michio Jimbo},
  title     = {Quantum {$R$} Matrix for the Generalized Toda System},
  journal   = {Communications in Mathematical Physics},
  volume    = {102},
  number    = {4},
  pages     = {537--547},
  year      = {1986},
  doi       = {10.1007/BF01221646},
  url       = {https://doi.org/10.1007/BF01221646}
}

@article{Lima_Santos_A1n,
  title         = {{$A^{(1)}_n$} reflection K-matrices},
  author        = {Lima-Santos, A.},
  journal       = {Nuclear Physics B},
  volume        = {644},
  number        = {3},
  pages         = {568--584},
  year          = {2002},
  month         = nov,
  publisher     = {Elsevier BV},
  doi           = {10.1016/S0550-3213(02)00819-2},
  issn          = {0550-3213},
  archivePrefix = {arXiv},
  eprint        = {nlin/0207028}
}

@article{Lima_Santos_B1n_A22n,
   title={{$B^{(1)}_n$} and {$A^{(2)}_{2n}$} reflection K-matrices},
   volume={654},
   ISSN={0550-3213},
   url={http://dx.doi.org/10.1016/S0550-3213(03)00042-7},
   DOI={10.1016/s0550-3213(03)00042-7},
   number={3},
   journal={Nuclear Physics B},
   publisher={Elsevier BV},
   author={Lima-Santos, A.},
   year={2003},
   month=mar, pages={466–480},
  archivePrefix = {arXiv},
  eprint        = {nlin/0210046}

}

@article{Lima_Santos_C1n_D1n_A22n1,
   title={{$C^{(1)}_n$}, {$D^{(1)}_n$} and {$A^{(2)}_{2n-1}$} reflection K-matrices},
   volume={675},
   ISSN={0550-3213},
   url={http://dx.doi.org/10.1016/j.nuclphysb.2003.09.037},
   DOI={10.1016/j.nuclphysb.2003.09.037},
   number={3},
   journal={Nuclear Physics B},
   publisher={Elsevier BV},
   author={Lima-Santos, A. and Malara, R.},
   year={2003},
   month=dec, pages={661–684},
  archivePrefix = {arXiv},
  eprint        = {nlin/0307046}

}

@article{Lima_Santos_D2n,
   title={{$D_{n+1}^{(2)}$} reflection K-matrices},
   volume={612},
   ISSN={0550-3213},
   url={http://dx.doi.org/10.1016/S0550-3213(01)00345-5},
   DOI={10.1016/s0550-3213(01)00345-5},
   number={3},
   journal={Nuclear Physics B},
   publisher={Elsevier BV},
   author={Lima-Santos, A.},
   year={2001},
   month=oct, pages={446–460},
   archivePrefix = {arXiv},
   eprint={nlin/0104062}
}

@article{Lima_Santos_All,
   title={On {$A^{(1)}_n$}, {$B^{(1)}_n$}, {$C^{(1)}_n$}, {$D^{(1)}_n$}, {$A^{(2)}_{2n}$}, {$A^{(2)}_{2n-1}$} and {$D^{(2)}_{n+1}$} reflection K-matrices},
   volume={2006},
   ISSN={1742-5468},
   url={http://dx.doi.org/10.1088/1742-5468/2006/09/P09013},
   DOI={10.1088/1742-5468/2006/09/p09013},
   number={09},
   journal={Journal of Statistical Mechanics: Theory and Experiment},
   publisher={IOP Publishing},
   author={Malara, R and Lima-Santos, A},
   year={2006},
   month=sep, pages={P09013–P09013},
   archivePrefix = {arXiv},
   eprint={nlin/0412058}
}

@article{Sklyanin_1988,
doi = {10.1088/0305-4470/21/10/015},
url = {https://doi.org/10.1088/0305-4470/21/10/015},
year = {1988},
month = {may},
publisher = {},
volume = {21},
number = {10},
pages = {2375},
author = {E K Sklyanin},
title = {Boundary conditions for integrable quantum systems},
journal = {Journal of Physics A: Mathematical and General}
}

@article{Mezincescu:1990hda,
    author = "Mezincescu, Luca and Nepomechie, Rafael I.",
    title = "{Integrability of open spin chains with quantum algebra symmetry}",
    eprint = "hep-th/9206047",
    archivePrefix = "arXiv",
    reportNumber = "UMTG-159",
    doi = "10.1142/S0217751X9200257X",
    journal = "Int. J. Mod. Phys. A",
    volume = "6",
    pages = "5231--5248",
    year = "1991",
    note = "[Addendum: Int. J. Mod. Phys. A 7, 5657--5659 (1992)]"
}

@article{Caux_2011,
doi = {10.1088/1742-5468/2011/02/P02023},
url = {https://doi.org/10.1088/1742-5468/2011/02/P02023},
year = {2011},
month = {feb},
publisher = {},
volume = {2011},
number = {02},
pages = {P02023},
author = {Caux, Jean-Sébastien and Mossel, Jorn},
title = {Remarks on the notion of quantum integrability},
journal = {Journal of Statistical Mechanics: Theory and Experiment},

eprint={arXiv:1012.3587}
}

@article{Vernier24,
  title = {Strong Zero Modes in Integrable Quantum Circuits},
  author = {Vernier, Eric and Yeh, Hsiu-Chung and Piroli, Lorenzo and Mitra, Aditi},
  journal = {Phys. Rev. Lett.},
  volume = {133},
  issue = {5},
  pages = {050606},
  numpages = {7},
  year = {2024},
  month = {Aug},
  publisher = {American Physical Society},
  doi = {10.1103/PhysRevLett.133.050606},
  eprint={arXiv:2401.12305}
}

@misc{Fendley2025,
      title={{XYZ} integrability the easy way}, 
      author={Paul Fendley and Sascha Gehrmann and Eric Vernier and Frank Verstraete},
      year={2025},
      eprint={2511.04674},
      archivePrefix={arXiv},
      primaryClass={cond-mat.stat-mech},
      url={https://arxiv.org/abs/2511.04674}, 
}

@misc{Gehrmann2025,
      title={Exact strong zero modes in quantum circuits and spin chains with non-diagonal boundary conditions}, 
      author={Sascha Gehrmann and Fabian H. L. Essler},
      year={2025},
      eprint={2511.05490},
      archivePrefix={arXiv},
      primaryClass={cond-mat.stat-mech},
      url={https://arxiv.org/abs/2511.05490}, 
}

@misc{Essler2025,
      title={Strong zero modes in integrable spin-S chains}, 
      author={Fabian H. L. Essler and Paul Fendley and Eric Vernier},
      year={2025},
      eprint={2512.07742},
      archivePrefix={arXiv},
      primaryClass={cond-mat.stat-mech},
      url={https://arxiv.org/abs/2512.07742}, 
}

@misc{Moudgalya2026,
      title={Strong Zero Modes via Commutant Algebras}, 
      author={Sanjay Moudgalya and Olexei I. Motrunich},
      year={2026},
      eprint={2603.02326},
      archivePrefix={arXiv},
      primaryClass={cond-mat.stat-mech},
      url={https://arxiv.org/abs/2603.02326}, 
}

@article{Fendley_2016,
doi = {10.1088/1751-8113/49/30/30LT01},
url = {https://doi.org/10.1088/1751-8113/49/30/30LT01},
year = {2016},
month = {jun},
publisher = {IOP Publishing},
volume = {49},
number = {30},
pages = {30LT01},
author = {Fendley, Paul},
title = {Strong zero modes and eigenstate phase transitions in the XYZ/interacting Majorana chain},
journal = {Journal of Physics A: Mathematical and Theoretical}
}

@article{mukherjee2024emergent,
  title = {Emergent strong zero mode through local Floquet engineering},
  author = {Mukherjee, Bhaskar and Melendrez, Ronald and Szyniszewski, Marcin and Changlani, Hitesh J. and Pal, Arijeet},
  journal = {Phys. Rev. B},
  volume = {109},
  issue = {6},
  pages = {064303},
  numpages = {23},
  year = {2024},
  month = {Feb},
  publisher = {American Physical Society},
  doi = {/10.1103/PhysRevB.109.064303}
}

@article{Potter18,
  title = {String order parameters for one-dimensional Floquet symmetry protected topological phases},
  author = {Kumar, Ajesh and Dumitrescu, Philipp T. and Potter, Andrew C.},
  journal = {Phys. Rev. B},
  volume = {97},
  issue = {22},
  pages = {224302},
  numpages = {9},
  year = {2018},
  month = {Jun},
  publisher = {American Physical Society},
  doi = {10.1103/PhysRevB.97.224302},
  url = {https://link.aps.org/doi/10.1103/PhysRevB.97.224302}
}

@article{Yao17,
  title = {Floquet Symmetry-Protected Topological Phases in Cold-Atom Systems},
  author = {Potirniche, I.-D. and Potter, A. C. and Schleier-Smith, M. and Vishwanath, A. and Yao, N. Y.},
  journal = {Phys. Rev. Lett.},
  volume = {119},
  issue = {12},
  pages = {123601},
  numpages = {6},
  year = {2017},
  month = {Sep},
  publisher = {American Physical Society},
  doi = {10.1103/PhysRevLett.119.123601},
  url = {https://link.aps.org/doi/10.1103/PhysRevLett.119.123601}
}

@article{sreejith2016parafermion,
  title = {Parafermion chain with $2\ensuremath{\pi}/k$ Floquet edge modes},
  author = {Sreejith, G. J. and Lazarides, Achilleas and Moessner, Roderich},
  journal = {Phys. Rev. B},
  volume = {94},
  issue = {4},
  pages = {045127},
  numpages = {9},
  year = {2016},
  month = {Jul},
  publisher = {American Physical Society},
  doi = {10.1103/PhysRevB.94.045127},
  url = {https://link.aps.org/doi/10.1103/PhysRevB.94.045127}
}

@article{Khemani16,
  title = {Phase Structure of Driven Quantum Systems},
  author = {Khemani, Vedika and Lazarides, Achilleas and Moessner, Roderich and Sondhi, S. L.},
  journal = {Phys. Rev. Lett.},
  volume = {116},
  issue = {25},
  pages = {250401},
  numpages = {6},
  year = {2016},
  month = {Jun},
  publisher = {American Physical Society},
  doi = {10.1103/PhysRevLett.116.250401},
  url = {https://link.aps.org/doi/10.1103/PhysRevLett.116.250401}
}

@article{iadecola2015stroboscopic,
  title = {Stroboscopic symmetry-protected topological phases},
  author = {Iadecola, Thomas and Santos, Luiz H. and Chamon, Claudio},
  journal = {Phys. Rev. B},
  volume = {92},
  issue = {12},
  pages = {125107},
  numpages = {9},
  year = {2015},
  month = {Sep},
  publisher = {American Physical Society},
  doi = {10.1103/PhysRevB.92.125107}
}

@article{Sen13,
  title = {Floquet generation of Majorana end modes and topological invariants},
  author = {Thakurathi, Manisha and Patel, Aavishkar A. and Sen, Diptiman and Dutta, Amit},
  journal = {Phys. Rev. B},
  volume = {88},
  issue = {15},
  pages = {155133},
  numpages = {13},
  year = {2013},
  month = {Oct},
  publisher = {American Physical Society},
  doi = {10.1103/PhysRevB.88.155133}
}

@article{Bahri15,
    author = {Y. Bahri and R. Ronen and E. Altman},
    journal = {Nature Communications},
    title = {Localization and topology protected quantum coherence at the edge of hot matter},
    volume = {6},
    pages = {7341},
    year = {2015},
    doi={10.1038/ncomms8341}
}

@misc{Prosen:2026knv,
    author = "Prosen, Toma{\v{z}}",
    title = "{Quasi-local Edge Mode in XXX Spin Chain/Circuit with Interaction Boundary Defect}",
    eprint = "2603.17835",
    archivePrefix = "arXiv",
    primaryClass = "cond-mat.stat-mech",
    month = "3",
    year = "2026"
}

@article{Alicea_2012,
doi = {10.1088/0034-4885/75/7/076501},
url = {https://doi.org/10.1088/0034-4885/75/7/076501},
year = {2012},
month = {jun},
publisher = {IOP Publishing},
volume = {75},
number = {7},
pages = {076501},
author = {Alicea, Jason},
title = {New directions in the pursuit of Majorana fermions in solid state systems},
journal = {Reports on Progress in Physics}
}

@article{scienceKickedIsing,
author = {X. Mi  and M. Sonner  and M. Y. Niu  and K. W. Lee  and B. Foxen  and R. Acharya  and I. Aleiner  and T. I. Andersen  and F. Arute  and K. Arya  and et al },
title = {Noise-resilient edge modes on a chain of superconducting qubits},
journal = {Science},
volume = {378},
number = {6621},
pages = {785-790},
year = {2022},
doi = {10.1126/science.abq5769},
URL = {https://www.science.org/doi/abs/10.1126/science.abq5769},
eprint = {https://www.science.org/doi/pdf/10.1126/science.abq5769}
}

@article{yates2019almost,
  title = {Almost strong $(0,\ensuremath{\pi})$ edge modes in clean interacting one-dimensional Floquet systems},
  author = {Yates, Daniel J. and Essler, Fabian H. L. and Mitra, Aditi},
  journal = {Phys. Rev. B},
  volume = {99},
  issue = {20},
  pages = {205419},
  numpages = {13},
  year = {2019},
  month = {May},
  publisher = {American Physical Society},
  doi = {10.1103/PhysRevB.99.205419}
}

@misc{note1,
  author="{An example of a R-matrix without crossing symmetry is the $A^{(1)}_{n-1}$ series for $n>2$}",
  title = "{however, still an appropriate  $M$ to consider open BCs can be defined see \cite{Reshetikhin1990, Mezincescu:1990hda}.}"
}

@misc{note2, 
author="{One can also work in a setting where $R$ is not periodic but quasi-periodic}",
title="{yielding the identity $K(\tfrac{p}{2})\propto U$. Then we still obtain a pull-though-identity by $R_{12}(p/2-v)K_1(p/2)R_{12}(p/2+v)\propto R_{12}(p/2-v)U_1R_{12}(p/2+v)\propto R_{12}(p/2-v)R_{12}(-p/2+v)U_{1}\propto U_1$. An example of this is indeed the gauge used in \cite{Vernier24}. Here, $R$ is quasi-periodic with $p=\ri \pi$ and $U=\sigma^z$.}" 
}

@misc{Supplemental, 
note="{See Supplemental Material for technical details.}" 
}

@article{Nienhuis1982,
  author    = {Nienhuis, B.},
  title     = {Exact critical point and critical exponents of {$O(n)$} models in two dimensions},
  journal   = {Physical Review Letters},
  volume    = {49},
  number    = {15},
  pages     = {1062--1065},
  year      = {1982},
  doi       = {10.1103/PhysRevLett.49.1062},
}

@article{Nienhuis1990,
  author    = {Nienhuis, B.},
  title     = {Critical spin-1 vertex models and {$O(n)$} models},
  journal   = {International Journal of Modern Physics B},
  volume    = {4},
  number    = {5},
  pages     = {929--942},
  year      = {1990},
  doi       = {10.1142/S0217979290000449},
}

@article{Yung1995,
  author    = {Yung, C. M. and Batchelor, M. T.},
  title     = {Integrable vertex and loop models on the square lattice with open boundaries via reflection matrices},
  journal   = {Nuclear Physics B},
  volume    = {435},
  number    = {3},
  pages     = {430--462},
  year      = {1995},
  doi       = {10.1016/0550-3213(94)00448-N},
  eprint    = {hep-th/9410042},
  archivePrefix = {arXiv},
}

@article{Vernier2014,
  author    = {Vernier, Eric and Jacobsen, Jesper Lykke and Saleur, Hubert},
  title     = {Non-compact conformal field theory and the {$a_2^{(2)}$} ({Izergin--Korepin}) model in regime {III}},
  journal   = {Journal of Physics A: Mathematical and Theoretical},
  volume    = {47},
  number    = {28},
  pages     = {285202},
  year      = {2014},
  doi       = {10.1088/1751-8113/47/28/285202},
  eprint    = {1404.4497},
  archivePrefix = {arXiv},
}

@article{Vernier2015,
  author    = {Vernier, Eric and Jacobsen, Jesper Lykke and Saleur, Hubert},
  title     = {A new look at the collapse of two-dimensional polymers},
  journal   = {Journal of Statistical Mechanics: Theory and Experiment},
  volume    = {2015},
  pages     = {P09001},
  year      = {2015},
  doi       = {10.1088/1742-5468/2015/09/P09001},
  eprint    = {1505.07007},
  archivePrefix = {arXiv},
}

@book{KorepinBogoliubovIzergin1993,
  author    = {Vladimir E. Korepin and Nikolai M. Bogoliubov and Anatoli G. Izergin},
  title     = {Quantum Inverse Scattering Method and Correlation Functions},
  series    = {Cambridge Monographs on Mathematical Physics},
  publisher = {Cambridge University Press},
  address   = {Cambridge},
  year      = {1993},
  isbn      = {9780521373203}
}

@article{Yang1968,
  author    = {Yang, C. N.},
  title     = {{$S$} matrix for the one-dimensional {$N$}-body problem
               with repulsive or attractive $\delta$-function interaction},
  journal   = {Physical Review},
  volume    = {168},
  number    = {5},
  pages     = {1920--1923},
  year      = {1968},
  doi       = {10.1103/PhysRev.168.1920},
}

@article{Baxter1972,
  author    = {Baxter, R. J.},
  title     = {Partition function of the eight-vertex lattice model},
  journal   = {Annals of Physics},
  volume    = {70},
  number    = {1},
  pages     = {193--228},
  year      = {1972},
  doi       = {10.1016/0003-4916(72)90335-1},
}

@book{Baxter1982,
  author    = {Baxter, R. J.},
  title     = {Exactly Solved Models in Statistical Mechanics},
  publisher = {Academic Press},
  address   = {London},
  year      = {1982},
}

@article{Cherednik1984,
  author       = {Ivan V. Cherednik},
  title        = {Factorizing Particles on a Half-Line and Root Systems},
  journal      = {Theoretical and Mathematical Physics},
  year         = {1984},
  volume       = {61},
  number       = {1},
  pages        = {977--983},
  doi          = {10.1007/BF01038545},
}

@article{FaddeevTakhtajan1979,
  author       = {L. D. Faddeev and L. A. Takhtajan},
  title        = {The Quantum Method of the Inverse Problem and the {H}eisenberg {XYZ} Model},
  journal      = {Russian Mathematical Surveys},
  year         = {1979},
  volume       = {34},
  number       = {5},
  pages        = {11--68},
  doi          = {10.1070/RM1979v034n05ABEH003909},
}

@misc{joshi2026tunablefloquetselectionrules,
      title={Tunable Floquet selection rules in a driven Ising chain}, 
      author={Rishi Paresh Joshi and Sanchayan Banerjee and Sneha Narasimha Moorthy and Tapan Mishra},
      year={2026},
      eprint={2603.23493},
      archivePrefix={arXiv},
      primaryClass={cond-mat.other},
      url={https://arxiv.org/abs/2603.23493}, 
}

@misc{kantha2026strongzeromodesrandom,
      title={Strong zero modes in random Ising-Majorana chains}, 
      author={Saurav Kantha and Nicolas Laflorencie},
      year={2026},
      eprint={2603.05313},
      archivePrefix={arXiv},
      primaryClass={cond-mat.dis-nn},
      url={https://arxiv.org/abs/2603.05313}, 
}

@article{PhysRevB.88.014206,
  title = {Localization-protected quantum order},
  author = {Huse, David A. and Nandkishore, Rahul and Oganesyan, Vadim and Pal, Arijeet and Sondhi, S. L.},
  journal = {Phys. Rev. B},
  volume = {88},
  issue = {1},
  pages = {014206},
  numpages = {8},
  year = {2013},
  month = {Jul},
  publisher = {American Physical Society},
  doi = {10.1103/PhysRevB.88.014206},
  url = {https://link.aps.org/doi/10.1103/PhysRevB.88.014206}
}

@misc{katz2025,
      title={Hybrid digital-analog protocols for simulating quantum multi-body interactions}, 
      author={Or Katz and Alexander Schuckert and Tianyi Wang and Eleanor Crane and Alexey V. Gorshkov and Marko Cetina},
      year={2025},
      eprint={2512.21385},
      archivePrefix={arXiv},
      primaryClass={quant-ph},
      url={https://arxiv.org/abs/2512.21385}, 
}

@article{Datla_2026,
   title={Statistical localization of U(1) lattice gauge theory in a Rydberg simulator},
   volume={22},
   ISSN={1745-2481},
   url={http://dx.doi.org/10.1038/s41567-026-03183-w},
   DOI={10.1038/s41567-026-03183-w},
   number={3},
   journal={Nature Physics},
   publisher={Springer Science and Business Media LLC},
   author={Datla, Prithvi Raj and Zhao, Luheng and Ho, Wen Wei and Klco, Natalie and Loh, Huanqian},
   year={2026},
   month=Feb, pages={355–361} }

@article{Kuno:2023ass,
    author = "Kuno, Yoshihito and Ichinose, Ikuo",
    title = "{Interplay between lattice gauge theory and subsystem codes}",
    eprint = "2304.05718",
    archivePrefix = "arXiv",
    primaryClass = "cond-mat.stat-mech",
    doi = "10.1103/PhysRevB.108.045150",
    journal = "Phys. Rev. B",
    volume = "108",
    number = "4",
    pages = "045150",
    year = "2023"
}

@Article{10.21468/SciPostPhys.14.6.140,
	title={{Fragmentation-induced localization and boundary charges in dimensions two and above}},
	author={Julius Lehmann and Pablo Sala de Torres-Solanot and Frank Pollmann and Tibor Rakovszky},
	journal={SciPost Phys.},
	volume={14},
	pages={140},
	year={2023},
	publisher={SciPost},
	doi={10.21468/SciPostPhys.14.6.140},
	url={https://scipost.org/10.21468/SciPostPhys.14.6.140},
}

@article{Moran_2017,
   title={Parafermionic clock models and quantum resonance},
   volume={95},
   ISSN={2469-9969},
   url={http://dx.doi.org/10.1103/PhysRevB.95.235127},
   DOI={10.1103/physrevb.95.235127},
   number={23},
   journal={Physical Review B},
   publisher={American Physical Society (APS)},
   author={Moran, N. and Pellegrino, D. and Slingerland, J. K. and Kells, G.},
   year={2017},
   month=June }

@article{PhysRevB.97.064424,
  title = {Infinite coherence time of edge spins in finite-length chains},
  author = {Maceira, Ivo A. and Mila, Fr\'ed\'eric},
  journal = {Phys. Rev. B},
  volume = {97},
  issue = {6},
  pages = {064424},
  numpages = {7},
  year = {2018},
  month = {Feb},
  publisher = {American Physical Society},
  doi = {10.1103/PhysRevB.97.064424},
  url = {https://link.aps.org/doi/10.1103/PhysRevB.97.064424}
}

@article{PhysRevB.98.094308,
  title = {Enhancing correlation times for edge spins through dissipation},
  author = {Vasiloiu, Loredana M. and Carollo, Federico and Garrahan, Juan P.},
  journal = {Phys. Rev. B},
  volume = {98},
  issue = {9},
  pages = {094308},
  numpages = {6},
  year = {2018},
  month = {Sep},
  publisher = {American Physical Society},
  doi = {10.1103/PhysRevB.98.094308},
  url = {https://link.aps.org/doi/10.1103/PhysRevB.98.094308}
}

@article{McGinley_2019,
   title={Slow Growth of Out-of-Time-Order Correlators and Entanglement Entropy in Integrable Disordered Systems},
   volume={122},
   ISSN={1079-7114},
   url={http://dx.doi.org/10.1103/PhysRevLett.122.020603},
   DOI={10.1103/physrevlett.122.020603},
   number={2},
   journal={Physical Review Letters},
   publisher={American Physical Society (APS)},
   author={McGinley, Max and Nunnenkamp, Andreas and Knolle, Johannes},
   year={2019},
   month=Jan }

@article{PhysRevB.101.104415,
  title = {Topologically induced prescrambling and dynamical detection of topological phase transitions at infinite temperature},
  author = {Da\ifmmode \breve{g}\else \u{g}\fi{}, Ceren B. and Duan, L.-M. and Sun, Kai},
  journal = {Phys. Rev. B},
  volume = {101},
  issue = {10},
  pages = {104415},
  numpages = {16},
  year = {2020},
  month = {Mar},
  publisher = {American Physical Society},
  doi = {10.1103/PhysRevB.101.104415},
  url = {https://link.aps.org/doi/10.1103/PhysRevB.101.104415}
}

@article{PhysRevResearch.4.L032016,
  title = {Topological order in random interacting Ising-Majorana chains stabilized by many-body localization},
  author = {Laflorencie, Nicolas and Lemari\'e, Gabriel and Mac\'e, Nicolas},
  journal = {Phys. Rev. Res.},
  volume = {4},
  issue = {3},
  pages = {L032016},
  numpages = {7},
  year = {2022},
  month = {Jul},
  publisher = {American Physical Society},
  doi = {10.1103/PhysRevResearch.4.L032016},
  url = {https://link.aps.org/doi/10.1103/PhysRevResearch.4.L032016}
}

@article{PhysRevB.101.125126,
  title = {Statistical localization: From strong fragmentation to strong edge modes},
  author = {Rakovszky, Tibor and Sala, Pablo and Verresen, Ruben and Knap, Michael and Pollmann, Frank},
  journal = {Phys. Rev. B},
  volume = {101},
  issue = {12},
  pages = {125126},
  numpages = {23},
  year = {2020},
  month = {Mar},
  publisher = {American Physical Society},
  doi = {10.1103/PhysRevB.101.125126},
  url = {https://link.aps.org/doi/10.1103/PhysRevB.101.125126}
}

@article{PRXQuantum.3.020330,
  title = {Symmetry-Protected Infinite-Temperature Quantum Memory from Subsystem Codes},
  author = {Wildeboer, Julia and Iadecola, Thomas and Williamson, Dominic J.},
  journal = {PRX Quantum},
  volume = {3},
  issue = {2},
  pages = {020330},
  numpages = {24},
  year = {2022},
  month = {May},
  publisher = {American Physical Society},
  doi = {10.1103/PRXQuantum.3.020330},
  url = {https://link.aps.org/doi/10.1103/PRXQuantum.3.020330}
}

@article{PhysRevB.107.224312,
  title = {Numerical methods for detecting symmetries and commutant algebras},
  author = {Moudgalya, Sanjay and Motrunich, Olexei I.},
  journal = {Phys. Rev. B},
  volume = {107},
  issue = {22},
  pages = {224312},
  numpages = {19},
  year = {2023},
  month = {Jun},
  publisher = {American Physical Society},
  doi = {10.1103/PhysRevB.107.224312},
  url = {https://link.aps.org/doi/10.1103/PhysRevB.107.224312}
}

@article{PRXQuantum.5.040330,
  title = {Symmetries as Ground States of Local Superoperators: Hydrodynamic Implications},
  author = {Moudgalya, Sanjay and Motrunich, Olexei I.},
  journal = {PRX Quantum},
  volume = {5},
  issue = {4},
  pages = {040330},
  numpages = {41},
  year = {2024},
  month = {Nov},
  publisher = {American Physical Society},
  doi = {10.1103/PRXQuantum.5.040330},
  url = {https://link.aps.org/doi/10.1103/PRXQuantum.5.040330}
}

@article{MOUDGALYA2023169384,
title = {From symmetries to commutant algebras in standard Hamiltonians},
journal = {Annals of Physics},
volume = {455},
pages = {169384},
year = {2023},
issn = {0003-4916},
doi = {https://doi.org/10.1016/j.aop.2023.169384},
url = {https://www.sciencedirect.com/science/article/pii/S0003491623001707},
author = {Sanjay Moudgalya and Olexei I. Motrunich}
}

@article{Monthus:2018blb,
    author = "Monthus, C{\'e}cile",
    title = "{Even and odd normalized zero modes in random interacting Majorana models respecting the parity $P$ and the time-reversal-symmetry $T$}",
    eprint = "1803.01348",
    archivePrefix = "arXiv",
    primaryClass = "cond-mat.dis-nn",
    doi = "10.1088/1751-8121/aac4b0",
    journal = "J. Phys. A",
    volume = "51",
    number = "26",
    pages = "265303",
    year = "2018"
}

@article{Mahyaeh:2019sxb,
    author = "Mahyaeh, Iman and Ardonne, Eddy",
    title = "{Study of the phase diagram of the Kitaev-Hubbard chain}",
    eprint = "1911.03156",
    archivePrefix = "arXiv",
    primaryClass = "cond-mat.str-el",
    doi = "10.1103/PhysRevB.101.085125",
    journal = "Phys. Rev. B",
    volume = "101",
    number = "8",
    pages = "085125",
    year = "2020"
}

@article{Munk:2018oan,
    author = "Munk, Morten I. Kj{\ae}rgaard and Rasmussen, A. and Burrello, M.",
    title = "{Dyonic zero-energy modes}",
    eprint = "1807.09286",
    archivePrefix = "arXiv",
    primaryClass = "cond-mat.str-el",
    reportNumber = "NBI CMT QDEV 2018",
    doi = "10.1103/PhysRevB.98.245135",
    journal = "Phys. Rev. B",
    volume = "98",
    number = "24",
    pages = "245135",
    year = "2018"
}

@article{Svensson:2024djf,
    author = "Svensson, Viktor and Leijnse, Martin",
    title = "{Quantum dot based Kitaev chains: Majorana quality measures and scaling with increasing chain length}",
    eprint = "2407.09211",
    archivePrefix = "arXiv",
    primaryClass = "cond-mat.mes-hall",
    doi = "10.1103/PhysRevB.110.155436",
    journal = "Phys. Rev. B",
    volume = "110",
    number = "15",
    pages = "155436",
    year = "2024"
}

@article{Yates:2021asz,
    author = "Yates, Daniel J. and Mitra, Aditi",
    title = "{Strong and almost strong modes of Floquet spin chains in Krylov subspaces}",
    eprint = "2105.13246",
    archivePrefix = "arXiv",
    primaryClass = "cond-mat.str-el",
    doi = "10.1103/PhysRevB.104.195121",
    journal = "Phys. Rev. B",
    volume = "104",
    number = "19",
    pages = "195121",
    year = "2021"
}

@article{Yeh:2023cwb,
    author = "Yeh, Hsiu-Chung and Cardoso, Gabriel and Korneev, Leonid and Sels, Dries and Abanov, Alexander G. and Mitra, Aditi",
    title = "{Slowly decaying zero mode in a weakly nonintegrable boundary impurity model}",
    eprint = "2305.11325",
    archivePrefix = "arXiv",
    primaryClass = "cond-mat.str-el",
    doi = "10.1103/PhysRevB.108.165143",
    journal = "Phys. Rev. B",
    volume = "108",
    number = "16",
    pages = "165143",
    year = "2023"
}

@article{Alicea:2015hja,
    author = "Alicea, Jason and Fendley, Paul",
    title = "{Topological phases with parafermions: theory and blueprints}",
    eprint = "1504.02476",
    archivePrefix = "arXiv",
    primaryClass = "cond-mat.str-el",
    doi = "10.1146/annurev-conmatphys-031115-011336",
    journal = "Ann. Rev. Condensed Matter Phys.",
    volume = "7",
    pages = "119",
    year = "2016"
}

@article{Tausendpfund:2025pyy,
    author = "Tausendpfund, Niklas and Mitra, Aditi and Rizzi, Matteo",
    title = "{Almost strong zero modes at finite temperature}",
    eprint = "2501.11121",
    archivePrefix = "arXiv",
    primaryClass = "cond-mat.str-el",
    doi = "10.1103/PhysRevResearch.7.023245",
    journal = "Phys. Rev. Res.",
    volume = "7",
    number = "2",
    pages = "023245",
    year = "2025"
}

@article{Kawabata:2017zsb,
    author = "Kawabata, Kohei and Kobayashi, Ryohei and Wu, Ning and Katsura, Hosho",
    title = "{Exact zero modes in twisted Kitaev chains}",
    eprint = "1702.00197",
    archivePrefix = "arXiv",
    primaryClass = "cond-mat.mes-hall",
    doi = "10.1103/PhysRevB.95.195140",
    journal = "Phys. Rev. B",
    volume = "95",
    number = "19",
    pages = "195140",
    year = "2017"
}

@article{Izergin1981,
  author    = {A. G. Izergin and V. E. Korepin},
  title     = {The inverse scattering method approach to the quantum Shabat-Mikhailov model},
  journal   = {Communications in Mathematical Physics},
  volume    = {79},
  number    = {3},
  pages     = {303--316},
  year      = {1981},
  doi       = {10.1007/BF01208496},
  url       = {https://doi.org/10.1007/BF01208496},
  issn      = {1432-0916}
}

@misc{frassek2026intertwiningmarkovprocessesmatrix,
      title={Intertwining Markov Processes via Matrix Product Operators}, 
      author={Rouven Frassek and Jan de Gier and Jimin Li and Frank Verstraete},
      year={2026},
      eprint={2603.09928},
      archivePrefix={arXiv},
      primaryClass={math-ph},
      url={https://arxiv.org/abs/2603.09928}, 
}

@misc{rubio2026localcharacterizationglobaltensor,
      title={The local characterization of global tensor network eigenstates}, 
      author={José Garre Rubio and András Molnár and Norbert Schuch and Frank Verstraete},
      year={2026},
      eprint={2603.28349},
      archivePrefix={arXiv},
      primaryClass={quant-ph},
      url={https://arxiv.org/abs/2603.28349}, 
}

@article{Gehrmann:2026ypf,
    author = "Gehrmann, Sascha and Essler, Fabian H. L.",
    title = "{Exact Quantum Many-Body Scars by a generalized Matrix-Product Ansatz}",
    eprint = "2605.03020",
    archivefix = "arXiv",
    primaryClass = "quant-ph",
    month = "5",
    year = "2026",
    journal=""
}

@misc{note4,
  author="{These generalizations were obtained from the spin-1/2 chain by fusion to arbitrary spin $S$ in the physical space}",
  title = "{  while keeping the auxiallary space $S=1/2$ fixed. Note, that also the specific case of the $A^{(1)}_{3}$ model covered here has been previously constructed in \cite{Essler2025} corresponding to $S=3/2$, by fusion in both the physical and auxiliary spaces.}"
}

	\newpage
	\appendix
	\renewcommand{\theequation}{A\arabic{equation}}
	\onecolumngrid
	\newpage
	\begin{center}
		\ \vskip 0.2cm
		{\large\bf End Matter}
	\end{center}
	\twocolumngrid
	
	\paragraph{\texorpdfstring{$R$-matrices and $K$-matrices}{R-matrices}\label{RKMats}.--}
	In the following, I recall the trigonometric solutions of the Yang-Baxter equation based on the non-exceptional Lie algebras $A^{(1)}_{n-1}$, $A^{(2)}_{2n-1}$, $A^{(2)}_{2n}$, $B^{(1)}_{n}$, $C^{(1)}_{n}$, $D^{(1)}_{n}$ following the convention of \cite{Jimbo1986}. In addition to this, I recall the associated crossing parameters $\crp$, crossing matrices $M$, and prefactors $\xi$ used in the crossing unitarity
	\begin{equation}
		\begin{aligned}
			R_{1,2}(u)M_1R^{t_2}_{1,2}&(-u-2\crp)M^{-1}_1\\&=\xi^2a_1(u+\crp)a_1(-u-\crp)\id\,
		\end{aligned}
	\end{equation}
	and the generalized crossing symmetry \cite{Mezincescu:1990hda,Reshetikhin1990} for $A^{(1)}_{n-1}$ 
	\begin{align*}
		\Big(\Big(\Big(\Big(R_{12}(u)\Big)^{t_2}\Big)^{-1}\Big)^{t_2}\Big)^{-1}=M_2R_{12}(u+2\crp)M^{-1}_2\,.
	\end{align*}
	Here, $a_1(u)$ is a Boltzmann weight in the R-matrix given below.  
	It is useful to introduce some general notation. Recall that the local Hilbert space dimension is denoted by $N$. I denote by $E_{ij}$ the elementary $N$ by $N$ matrices. All solutions have a free parameter $q$. Further, primed indices are set to be 
	\begin{align*}
		j'=N-j+1\,.
	\end{align*}
	For each series, I list an appropriate integrable boundary condition that 
	$\bra{\mathcal{K}}$ is orthogonal to $\ket{\mathcal{R}_0}$ and so the ESZM is localized at the left boundary.  Note that the following are just some examples; there exist more general BCs. In the following, left K-matrix is taken to be
	\begin{align}\label{elsnghdlrmgldjshsh}
		K^+(u\,|\,\{\beta^+_j\})=(K^-(-u-\crp\,|\,\{\beta^+_j\}))^tM\,.
	\end{align}
	Also, it is useful to define the function
	\begin{align}\label{akfnjsnfjfn}
		f_{i,i}(u) = \beta_{i,i}\bigl(e^{u}-1\bigr) + 1\,.
	\end{align}
	\paragraph{The series of $A^{(1)}_{n-1}$.--}
	The local Hilbert space dimension is given by $N=n$. The R-matrix is given by
	\begin{equation}\label{fjdfnjdnfjdnfjdöööö}
		\begin{aligned}
			R(u) &=a_{1}(u)\sum E_{ii}\otimes E_{ii}+a_{2}(u)\sum_{i\neq
				j}E_{ii}\otimes E_{jj} \\&+a_{3}(u)\sum_{i<j}E_{ij}\otimes E_{ji}+a_{4}(u)\sum_{i>j}E_{ij}\otimes
			E_{ji},  
		\end{aligned}
	\end{equation}%
	the functions $a_j$ are given by 
	\begin{equation}\label{rjnjsndjsnjdsn}
		\begin{aligned}
			a_{1}(u)=&(\re^{u}-q^{2}),\qquad a_{2}(u)=q(\re^{u}-1),\\
			a_{3}(u)=&-(q^{2}-1),\quad\,\, a_{4}(u)=-\re^{u}(q^{2}-1).  
		\end{aligned}
	\end{equation}%
	The crossing parameter $\crp$ and matrix $M$ are given by 
	\begin{align}
		\crp=-n\log(q)\,,\qquad \qquad M_{i,j}=\delta_{i,j} \, q^{n+1-2i} \,.
	\end{align}
	The BCs, so the ESZMs is localized, can be taken to be
	\begin{align}\label{fndjnfjdnfjdnfjd1}
		K^{-}(u\,|&\,\beta^-_{1,1}) = f_{11}(u)\,E_{11}
		+ \re^{2u} f_{11}(-u)\,E_{jj}\\
		&+ \re^{2u} f_{11}(-u) \sum_{l=2}^{j-1} E_{ll}
		+ \re^{2u} f_{11}(-u) \sum_{l=j+1}^{n} E_{ll}.\notag
	\end{align}
	One finds that the orthogonality condition is obey if 
	\begin{align}\label{fndjnfjdnfjdnfjd2}
		\beta^+_{1,1}=\frac{q^{n-2}}{1+q^{n-2}}\,.
	\end{align}
	\paragraph{The series of $B^{(1)}_{n}$, $C^{(1)}_n$, $D^{(1)}_{n}$ and $A^{(2)}_{2n}$, $A^{(2)}_{2n-1}$.--}%
	The local Hilbert space dimension and the parameter $\xi$ used in the following are given by 
	\begin{equation*}
		(N,\xi)=
		\begin{cases}
			(2n+1,\; q^{2n-1})   & B^{(1)}_{n},\\[4pt]
			(2n,\;   q^{2n+2})   & C^{(1)}_{n},\\[4pt]
			(2n,\;   q^{2n-2})   & D^{(1)}_{n},\\[4pt]
			(2n+1,\; -q^{2n+1})  & A^{(2)}_{2n},\\[4pt]
			(2n,\;   -q^{2n})    & A^{(2)}_{2n-1}.
		\end{cases}
	\end{equation*}

	The R-matrix is given by 
	\begin{align}
		R(u)&= a_1(u)\sum_{i\neq i'} E_{ii}\otimes E_{ii}
		+ a_2(u)\sum_{i\neq j,j'} E_{ii}\otimes E_{jj}\nonumber\\
		&+ a_3(u)\sum_{\substack{i<j\\ i\neq j'}} E_{ij}\otimes E_{ji}
		+ a_4(u)\sum_{\substack{i>j\\ i\neq j'}} E_{ij}\otimes E_{ji}\nonumber\\
		&+ \sum_{i,j} a_{ij}(u)\, E_{ij}\otimes E_{i'j'} \,,\label{kfefmdnfdjflafndfnf}
	\end{align}
	where the Boltzmann weights are now given by
	\begin{align}
		a_1(u) =& \big(e^{u}-q^{2}\big)\big(e^{u}-\xi\big)\,,\quad a_2(u) = q\big(e^{u}-1\big)\big(e^{u}-\xi\big),\notag\\
		a_3(u) =& -\big(q^{2}-1\big)\big(e^{u}-\xi\big)\,,\quad
		a_4(u) = e^{u}\,a_3(u),\label{tjenjsnfjyll}
	\end{align}
	and $a_{i,j}$ is given for $(i=j,\ i\neq i')$ as
	\begin{align}
		a_{ij}(u)
		&=
		\big(q^{2}\re^{u}-\xi\big)\big(\re^{u}-1\big),
	\end{align}
	for $(i=j,\ i=i')$ as
	\begin{align}
		a_{ij}(u)
		&=
		q\big(\re^{u}-\xi\big)\big(\re^{u}-1\big)+(\xi-1)(q^{2}-1)\re^{u},
	\end{align}
	for $ (i<j)$ as
	\begin{align}
		a_{ij}(u)
		&=
		(q^{2}-1)\big(\varepsilon_i\varepsilon_j\,\xi\,q^{\bar\imath-\bar\jmath}\,(\re^{u}-1)-\delta_{ij'}(\re^{u}-\xi)\big),
	\end{align}
	and for $(i>j)$ as 
	\begin{align}\label{kgdfkdnfjdnfjdnjfdn}
		a_{ij}(u)
		&=
		(q^{2}-1)\re^{u}\big(\varepsilon_i\varepsilon_j\,q^{\bar\imath-\bar\jmath}\,(\re^{u}-1)-\delta_{ij'}(\re^{u}-\xi)\big)\,.
	\end{align}
	Here for $B^{(1)}_{n}, D^{(1)}_{n}, A^{(2)}_{2n}, A^{(2)}_{2n-1}$ one has
	\begin{equation*}
		\begin{aligned}
			\bar{\imath} &=
			\begin{cases}
				i + \dfrac{1}{2}, & 1 \le i < \dfrac{N+1}{2},\\[6pt]
				i,               & i = \dfrac{N+1}{2},\\[6pt]
				i - \dfrac{1}{2}, & \dfrac{N+1}{2} < i \le N,
			\end{cases}
			\qquad
			\varepsilon_i = 1\,,
		\end{aligned}
	\end{equation*}
	while for $C^{(1)}_n$ one has 
	\begin{equation*}
		\begin{aligned}
			\bar{\imath} &=
			\begin{cases}
				i - \dfrac{1}{2}, & 1 \le i \le n,\\[6pt]
				i + \dfrac{1}{2}, & n< i \le 2n,
			\end{cases}\quad 
			\varepsilon_i &=
			\begin{cases}
				1,  & 1 \le i \le n,\\
				-1,  & n< i \le 2n.
			\end{cases}
		\end{aligned}
	\end{equation*}
	The crossing parameter $\crp$ and the matrix $M$ are
	\begin{align*}
		(\crp, M_{i,j}) =
		\begin{cases}
			\big(-\log(q^{2n-1}),\; \delta_{i,j}\, q^{2n+1-2\bar{i}}\big)
			& B^{(1)}_{n},\\[4pt]
			\big(-\log(q^{2n+1}),\; \delta_{i,j}\, q^{2n+1-2\bar{i}}\big)
			& C^{(1)}_{n},\\[4pt]
			\big(-\log(q^{2n-2}),\; \delta_{i,j}\, q^{2n+1-2\bar{i}}\big)
			& D^{(1)}_{n},\\[4pt]
			\big(-\log(-q^{2n+1}),\; \delta_{i,j}\, q^{2n+2-2\bar{i}}\big)
			& A^{(2)}_{2n},\\[4pt]
			\big(-\log(-q^{2n}),\; \delta_{i,j}\, q^{2n+1-2\bar{i}}\big)
			& A^{(2)}_{2n-1}.
		\end{cases}
	\end{align*}
	Now I report on appropriate BCs for the ESZM to be localized. For $B^{(1)}_n$ one can choose for instance, the diagonal matrix 
	\begin{equation}\label{skdskdnsjjf}
		\begin{aligned}
			K^-&({u \mid \beta^-_{1,1}}) = \frac{\beta^-_{1,1}\bigl(\re^{-u}-1\bigr)+2}{\beta^-_{1,1}\bigl(\re^{u}-1\bigr)+2}E_{11}+\sum^{2n}_{j=2}E_{jj}\\&+\frac{
				\beta^-_{1,1}\!\left(q^{2n-3}\re^{u}-1\right)+2}{
				\beta^-_{1,1}\!\left(q^{2n-3}\re^{-u}-1\right)+2
			}E_{2n+1,2n+1}
		\end{aligned}
	\end{equation}
	where for localization in the appropriate $\Delta$-regime one has the constraint:
	\begin{align}
		\beta^+_{1,1}=1\,, \qquad \text{or}\qquad \beta^+_{1,1}=\frac{2}{1+q^{2n-3}}\,.
	\end{align}
	
	For $C^{(1)}_n$ one can choose for instance
	\begin{align}\label{kejdhfkdoshhd}
		K^-({u \mid \beta^-_{1,1}}) = \sum^{n}_{j=1}E_{jj}+\frac{\beta^-_{1,1}\big(\re^{u}-1\big)-2}
		{\beta^-_{1,1}\big(\re^{-u}-1\big)-2}\sum^{2n}_{j=n+1}E_{jj}
	\end{align}
	and one need to set $\beta^+_{1,1}=-1$ for localization.
	For $D^{(1)}_2$ one can choose
	\begin{align}\label{kejdhfkdjdjdjoshhd1}
		&K^-(u\,|\,\{\beta^-_{1,1},\,\beta^-_{2,2}\})
		= E_{11}+ \frac{\beta^-_{2,2}(\re^{u}-1)-2}{\beta^-_{2,2}(\re^{-u}-1)-2}\,E_{33}\notag
		\\
		&+ \frac{\beta^-_{1,1}(\re^{u}-1)-2}{\beta^-_{1,1}(\re^{-u}-1)-2}\,\bigg(E_{22}+\frac{\beta^-_{2,2}(\re^{u}-1)-2}{\beta^-_{2,2}(\re^{-u}-1)-2}E_{44}\bigg)
	\end{align}
	while $D^{(1)}_n$ with $n>2$ one can choose for instance
	\begin{align}\label{kejdhfkdjdjdjoshhd2}
		&K^-({u \mid \beta^-_{1,1}}) = E_{11}+\frac{\beta^-_{1,1}\big(\re^{u}-1\big)-2}
		{\beta^-_{1,1}\big(\re^{-u}-1\big)-2}\sum^{2n-1}_{j=2}E_{jj}\\&+\frac{\big(\beta^-_{1,1}(\re^{u}-1)-2\big)\big(\beta^-_{1,1}(q^{2n-4}\re^{u}-1)-2\big)}
		{\big(\beta^-_{1,1}(\re^{-u}-1)-2\big)\big(\beta^-_{1,1}(q^{2n-4}\re^{-u}-1)-2\big)}E_{2n,2n}\,.\notag
	\end{align}
	In both cases, one can set $\beta^+_{1,1}=-1\,$ for localization.
	For $A^{(2)}_{2n-1}$ one can choose for instance
	\begin{align}
		&K^-({u \mid \beta^-_{1,1}}) = \sum^{n-1}_{j=1}E_{jj}+\frac{\beta^-_{1,1}\big(\re^{u}-1\big)-2}
		{\beta^-_{1,1}\big(\re^{-u}-1\big)-2}E_{nn}\label{eoksnfnfhhs}\\&+
		\re^{2u}\,
		\frac{\beta^-_{1,1}(\re^{-u}+q^{2n-2}) + 2q^{2n-2}}
		{\beta^-_{1,1}(\re^{u}+q^{2n-2}) + 2q^{2n-2}}E_{n+1,n+1}+
		\sum^{2n}_{j=n+2}\re^{2u}E_{jj}\notag
	\end{align}
	and one possible solution of the orthogonality condition is again $\beta^+_{1,1}=-1$.
	For $A^{(2)}_{2n}$ one can choose for instance
	\begin{align}
		K^-(&{u \mid \{\beta^-_{i,j}\}}) = 
		\sum_{i=1}^{n} \Big[1 + \beta^-_{1,1}(\re^{u}-1)\Big] E_{ii}\notag
		\\& + \left[
		\beta^-_{1,1}\re^{u}
		- \frac{\re^{2u}-q}{1-q}(\beta^-_{1,1}-1)
		\right] E_{n+1,n+1}\notag\\
		&+ \sum_{i=n+2}^{2n+1}
		\re^{2u}\Big[1 + \beta^-_{1,1}(\re^{-u}-1)\Big] E_{ii}
		\label{ksmdgnfjernenjd2n}\\&+ \sum_{i=1}^{n}
		\frac{1}{2}\beta^-_{i,i'}(\re^{2u}-1)\,E_{i,i'}\notag\\
		&+ \sum_{i=n+2}^{2n+1}
		\left(\frac{\beta^-_{1,1}-1}{q-1}\right)^2
		\frac{2q}{\beta^-_{i',i}}(\re^{2u}-1)\,E_{i,i'}\notag
	\end{align}
	and the solution of the orthogonality condition is
	$\beta^+_{1,1}=\frac{1}{2}$.
	
	The non-vanishing bulk Hamiltonian couplings for the IK chain \eqref{fjdnflsdmsffndjnfdj} corresponding to $n=1$ are given explicitly as 
	\begin{align}
		\lambda_{1,1}&=\lambda_{2,2}=\lambda_{6,6}=\lambda_{7,7}=2\Delta-1\,,\quad
		\lambda_{3,3}=\Delta\,,\notag\\
		\lambda_{4,4}&=\lambda_{5,5}=1  \,,\quad\lambda_{8,8}=\frac{1}{3}(4\Delta^2+\Delta-2)\notag\\
		\lambda_{6,1}&=\lambda_{7,2}= (\Delta-1+\sqrt{\Delta^2-1})(\Delta+\sqrt{\Delta^2-1})^{-\frac{3}{2}}\notag\\
		\lambda_{1,6}&=\lambda_{2,7}=-\lambda_{6,1}(\Delta+\sqrt{\Delta^2-1})^{2}\label{fjdnflsdmsffndjnfdjdsmmm}\\
		\lambda_{\substack{0,3\\3,0}}&=\frac{1}{3}(1\pm3\sqrt{\Delta-1}+\Delta(-5+4\Delta\mp 4\sqrt{\Delta^2-1}))\notag\\
		\lambda_{\substack{0,8\\8,0}}&=\frac{1}{3\sqrt{3}}(-1\pm9\sqrt{\Delta-1}+\Delta(5-4\Delta\mp 12\sqrt{\Delta^2-1}))\notag\\
		\lambda_{3,8}&=\frac{1}{\sqrt{3}}(-1+\Delta(-1+2\Delta+2\sqrt{\Delta^2-1}))\notag\\
		\lambda_{8,3}&=\frac{(1+\Delta-2\Delta^3-2\Delta^2\sqrt{\Delta^2-1})}{\sqrt{3}(\Delta+\sqrt{\Delta^2-1})^2}\notag\\
		J&=-\Big[2 \left(\sqrt{\Delta ^2-1}+\Delta -1\right) \sqrt{\sqrt{\Delta ^2-1}+\Delta }\Big]^{-1}\notag
	\end{align}

	\onecolumngrid
	\newpage 
	
	\begin{center}
		{\Large \bf Supplemental Material}
	\end{center}
	\setcounter{equation}{0}
	\setcounter{figure}{0}
	\renewcommand{\thetable}{S\arabic{table}}
	\renewcommand{\theequation}{S\thesection.\arabic{equation}}
	\renewcommand{\thefigure}{S\arabic{figure}}
	\setcounter{secnumdepth}{2}
	
	\section{Efficient computation of the Hilbert Schmidt Norms of \texorpdfstring{$\Psi_j$}{PSIj} and localisaztion condition}
	Here, I present the technical details underlying the efficient computation of the Hilbert–Schmidt (HS) norms used to diagnose the locality properties of the exact strong zero modes (ESZMs). At the end, the numerical results on the localization properties are presented in Fig.~S1.\\
	\textbf{Efficient computation of the HS norms:}	For convenience, recall that the Hilbert-Schmidt inner product of the two operators $A,B$ acting on a Hilbert Space $\mathscr{H}=\mathscr{V}^{\SS}$ of local dimension $N=\dim(\mathscr{V})$ induces the HS norm $|\!|A|\!|$ of an operator $A$:
	\begin{align}
		\langle A, B \rangle=\frac{1}{N^\SS}\,\tr(A^\dagger B ) \,,\qquad \qquad |\!|A|\!|^2=\frac{1}{N^\SS}\,\tr(A^\dagger A )\,.
	\end{align}
	One starts from the general definition of the ESZMs in the main text:
	\begin{align}\label{fdjfndjfnjdlslslsls}
		\mathbbG{\Psi}=\mathcal{N}\, \left(\mathbb{T}'\big(\tfrac{p}{2}\big)-\frac{\tr_{\mathcal{H}}\left(\mathbb{T}'\big(\tfrac{p}{2}\big)\right)}{{\rm dim}(\mathcal{H})}\,\id\right)=\mathcal{N}\left(\left(\sum^{\SS+1}_{j=1} \tilde{\Psi}_j\right)-\frac{\tr_{\mathcal{H}}\left(\mathbb{T}'\big(\tfrac{p}{2}\big)\right)}{{\rm dim}(\mathcal{H})}\,\id\right)\,,
	\end{align}
	here $\mathcal{N}$ is a yet to be determined scalar ensuring that $|\!|\mathbbG{\Psi}|\!|^2=1$ and the explicit form of the $\tilde{\Psi}_j$'s is given by 
	\begin{equation}\label{fdlaprejntejnjhs}
		\begin{aligned}
			\tilde{\Psi}_j=c_{\tilde{\Psi}_j}  \tr_{0}\bigg( &K^{+}_{0}(\tfrac{p}{2})\,\,R_{0,1}(\tfrac{p}{2})R_{0,2}(\tfrac{p}{2})\dots R_{0,j-1}(\tfrac{p}{2}) \left. \frac{{\rm d}}{{\rm d}u }\right|_{u=0}\,\Big[ R_{0,j}(u)\\
			&\times K^{-}_{0}(\tfrac{p}{2}) \,\,R_{j,0}(u)\Big] R_{j-1,0}(\tfrac{p}{2})\dots R_{2,0}(\tfrac{p}{2})R_{1,0}(\tfrac{p}{2})\bigg)\,,
			\\
			\tilde{\Psi}_{\SS+1}=  \tr_{0}\bigg( &K^{+}_{0}(\tfrac{p}{2})\,\,R_{0,1}(\tfrac{p}{2})\dots R_{0,\SS}(\tfrac{p}{2})\\
			&\times \left. \frac{{\rm d}}{{\rm d}u }\right|_{u=0}\, K^{-}_{0}(u) \,\,R_{\SS,0}(\tfrac{p}{2})\dots R_{1,0}(\tfrac{p}{2})
			\bigg)\,,
		\end{aligned}
	\end{equation}
	where the factor is given simply by
	\begin{align}
		c_{\tilde{\Psi}_j}=\left(\UniF(\tfrac{p}{2})\UniF(-\tfrac{p}{2})\right)^{\SS-j}\,.
	\end{align}
	Recall here that $a_1$ is given in the end matter for all considered R-matrices. To quantify the locality properties of an operator, a convenient approach is to consider its projection onto the part acting non-trivially on site $j$ and below:
	\begin{align}\label{wpkdncghtuslm3d}
		\Psi_j=\frac{1}{N^{L-j}}\tr_{j+1,\dots N} \left(\mathbbG{\Psi}\right)\otimes \id^{\SS-j-1}-\frac{1}{N^{L-j+1}}\tr_{j,\dots \SS} \left(\mathbbG{\Psi}\right)\otimes \id^{\SS-j}\,.
	\end{align}
	Further,  define $\tr_{\SS+1,\SS}(\mathbbG{\Psi})=\mathbbG{\Psi}$ such that 
	\begin{align}
		\Psi_{\SS}= \mathbbG{\Psi}-\frac{1}{N}\tr_{\SS} \left(\mathbbG{\Psi}\right)\otimes \id\,.
	\end{align}
	It follows trivially that
	\begin{align}
		\mathbbG{\Psi}=\sum^{\SS}_{j=1} \Psi_j\,.
	\end{align}
	Note that this is only valid as $\mathbbG{\Psi}$ has no contribution as the identity on the whole Hilbert space by definition \eqref{fdjfndjfnjdlslslsls}.
	Note, that the $\Psi_j$ are mutually orthogonal in the HS inner product.
	In the following, one seeks a form of $|\!|\Psi_j|\!|$ which is efficient for numerical evaluation. To start consider the expression \eqref{wpkdncghtuslm3d}, one has by \eqref{fdjfndjfnjdlslslsls}:
	\begin{align}\label{arjnwjnwjoloooo}
		\Psi_j=\mathcal{N}\cdot\sum^{\SS+1}_{k=j} \frac{1}{N^{L-j}}\tr_{j+1,\dots ,\SS}(\tilde{\Psi}_k)-\frac{1}{N^{L-j+1}}\tr_{j,\dots ,\SS}(\tilde{\Psi}_k)\,,
	\end{align}
	Note that, the identity term appearing in \eqref{fdjfndjfnjdlslslsls} cancelled out. 
	For efficient numerical evaluation via the Matrix-Product-Operator framework, one inserts \eqref{fdlaprejntejnjhs}, changes to index notation and groups the following combinations together
	\begin{align}
		W^{\mathbf{k}, \beta}_{\mathbf{j},\alpha}=& [R(\tfrac{p}{2})]^{k_1\beta}_{j_1\gamma}[R(\tfrac{p}{2})]^{\gamma j_2}_{\alpha k_2}\,,\quad\, \tilde{D}^{\mathbf{k}, \beta}_{\mathbf{j},\alpha}=\left.\frac{\rm d}{{\rm d}u}\right|_{u=\tfrac{p}{2}} [R(u)]^{k_1\beta}_{j_1\gamma}[R(u)]^{\gamma j_2}_{\alpha k_2}\,,\\ \big[\vec{K}^+\big]_{\mathbf{k}}=&[K^+(\tfrac{p}{2})]^{k_2}_{k_1}\qquad\qquad\qquad \vec{K}^-_{\mathbf{k}}=[K^-(\tfrac{p}{2})]^{k_1}_{k_2}\,,
	\end{align}
	where the bold indices are bi-indices of an enlarged "auxiliary space" by lexicographic order $\mathbf{k}=(k_1,k_2)$ and summing over repeated indices is implied. If one denotes by a bold zero $\mathbf{0}$ the enlarged auxiallary space, one can express the individual terms $\tilde{\Psi}_k$ in the following way: 
	\begin{align}\label{fff}
		\tilde{\Psi}_j&=c_{\tilde{\Psi}_j}\vec{K}^{+}_\mathbf{0} W_{\mathbf{0},1}W_{\mathbf{0},2}\dots W_{\mathbf{0},j-1} \tilde{D}_{\mathbf{0},j} \vec{K}^{-}_\mathbf{0}\,,\\
		\tilde{\Psi}_{\SS+1}&=\vec{K}^{+}_\mathbf{0} W_{\mathbf{0},1}W_{\mathbf{0},2}\dots W_{\mathbf{0},L} \vec{K}^{-'}_\mathbf{0}\,.
	\end{align}
	In the following, it is useful, for notational clarity, to represent the tensors graphically as
	\begin{center}
		\begin{tikzpicture}
			\node[above] at (-1.5-1.5,0.45) {$\vec{K}^{+} =$};
			\TenW{+}{-1+0.5-1.5}{0.5}{0.5}{0.5}{0}{0.25}{0}{0}
			
			\node[above] at (-1.5+2,0.5) {$\vec{K}^{-}=$};
			\TenW{-}{-1+0.5+2.25}{0.5}{0.5}{0.5}{0.25}{0}{0}{0}
			
			\node[above] at (-1.5+5.5,0.5) {$W=$};
			\TenW{W}{-1+0.5+5.5}{0.5}{0.5}{0.5}{0.25}{0.25}{0.25}{0.25}
			\node[above] at (6,0.45) {.};
		\end{tikzpicture}
	\end{center}
	In diagrammatic language $\tilde{\Psi}_j$ is given by 
	\begin{center}
		\begin{tikzpicture}
			\node[above] at (-1.5,0.45) {$\tilde{\Psi}_j=$};
			\TenW{+}{-1+0.25}{0.5}{0.5}{0.5}{0}{0.}{0}{0}
			\TenW{W}{0}{0.5}{0.5}{0.5}{0.25}{0.25}{0.25}{0.25}
			\TenW{W}{1}{0.5}{0.5}{0.5}{0.25}{0.25}{0.25}{0.25}
			\tikzDotsH{2.25}{0.75}
			\TenW{W}{3}{0.5}{0.5}{0.5}{0.25}{0.25}{0.25}{0.25}
			\TenW{\tilde{D}}{4}{0.5}{0.5}{0.5}{0.25}{0.25}{0.25}{0.25}
			\TenW{-}{5-0.25}{0.5}{0.5}{0.5}{0}{0}{0}{0}
			\node[above] at (5.5,0.45) {.};
		\end{tikzpicture}
	\end{center}
	Now by direct evaluation, one gets that \eqref{arjnwjnwjoloooo} can be rewritten as
	\begin{align}\label{gnjdngjngjng}
		\Psi_j=\mathcal{N}\cdot\,\vec{K}^{+}_\mathbf{0} W_{\mathbf{0},1}W_{\mathbf{0},2}\dots W_{\mathbf{0},j-1} \vec{D}_{\mathbf{0},j} 
	\end{align}
	with 
	\begin{equation}\label{ndjsjlsenensjsj}
		\begin{aligned}
			\vec{D}_{\mathbf{0},j}=\left( c_{\tilde{\Psi}_j}\tilde{D}^{\smallcancel{\id}}_{0,j} \vec{K}^{-}_0\right.&+\sum^\SS_{k=j+1} c_{\tilde{\Psi}_k} N^{j-k} W^{\smallcancel{\id}}_{\mathbf{0},j}\tr_{j+1}(W_{\mathbf{0},j+1})\dots \tr_{k-1}(W_{\mathbf{0},k-1}) \tr_{k}(\tilde{D}_{\mathbf{0},k}) \vec{K}^{-}_\mathbf{0}\\
			&+\left.N^{j-\SS}W^{\smallcancel{\mathbbm{1}}}_{0,j} \tr_{j+1}(W_{\mathbf{0},j+1})\dots \tr_{\SS}(W_{\mathbf{0},\SS})\vec{K}^{(-)'}_\mathbf{0}\right)
		\end{aligned}
	\end{equation}
	where in the final term the product $\tr(W_{\mathbf{0},j+1})\dots \tr(W_{\mathbf{0},L})$ is defined to be $\id$ for $j=L$ and the following notation has been introduced
	\begin{align}\label{ksdksnfjsnfjnfjeo}
		\mathcal{O}^{\smallcancel{\id}}_{0,j}:=\mathcal{O}_{0,j}-\frac{\tr_{j}(\mathcal{O}_{0,j})}{N}\id_j\, .
	\end{align}
	Note that the latter definition can be viewed as projecting onto the subspace orthogonal to the identity in the physical channel.
	The above steps can be expressed in diagrammatic language as 
	\begin{center}
		\begin{tikzpicture}
			\node[above] at (-1.75,0.4) {$\Psi_j=\mathcal{N}\cdot$};
			\TenW{+}{-1+0.25}{0.5}{0.5}{0.5}{0}{0}{0}{0}
			\TenW{W}{0}{0.5}{0.5}{0.5}{0.25}{0.25}{0.25}{0.25}
			\TenW{W}{1}{0.5}{0.5}{0.5}{0.25}{0.25}{0.25}{0.25}
			\tikzDotsH{2.25}{0.75}
			\TenW{W}{3}{0.5}{0.5}{0.5}{0.25}{0.25}{0.25}{0.25}
			\TenW{D_j}{4}{0.5}{0.5}{0.5}{0.25}{0}{0.25}{0.25}
			\node[above] at (4.75,0.45) {.};
		\end{tikzpicture}
	\end{center}
	Here the sum has been grouped into one tensor as in \eqref{ndjsjlsenensjsj}: 
	\begin{center}
		\begin{tikzpicture}
			\node[right] at (-4.15,0.725) {$=$};
			\TenW{D_{j}}{-5}{0.5}{0.5}{0.5}{0.25}{0}{0.25}{0.25}

			\node[] at (-3.,-0.25) {$j$};
			
			\node[right] at (-2,0.75) {$+\sum^\SS_{k=j+1}$};
			\TenW{D_{\id}}{-4+0.75}{0.5}{0.5}{0.5}{0.25}{0.25}{0.25}{0.25}
			\NoId{-4+0.75}{0.5}
			\TenW{-}{-3-0.25+0.75}{0.5}{0.5}{0.5}{0.25}{0}{0}{0}
			
			\TenW{W_{\id}}{0}{0.5}{0.5}{0.5}{0.25}{0.25}{0.25}{0.25}
			\NoId{0}{0.5}
			\TenW{W}{1}{0.5}{0.5}{0.5}{0.25}{0.25}{0.25}{0.25}
			\tikzDotsH{2.25}{0.75}
			\TenW{W}{3}{0.5}{0.5}{0.5}{0.25}{0.25}{0.25}{0.25}
			\TenW{D}{4}{0.5}{0.5}{0.5}{0.25}{0.25}{0.25}{0.25}
			\TenW{-}{5-0.25}{0.5}{0.5}{0.5}{0.25}{0}{0}{0}
			\trten{+1}{0.25}{+1}{0.75}
			\trten{+3}{0.25}{+3}{0.75}
			\trten{+4}{0.25}{+4}{0.75}
			\node[] at (0.25,-0.25) {$j$};
			\node[] at (1.25,-0.25) {$j+1$};
			\node[] at (3.25,-0.25) {$k-1$};
			\node[] at (4.25,-0.25) {$k$};
			
			\node[] at (5.5,0.75) {$+$};
			\def\x{6}

			\TenW{W_{\id}}{0+\x}{0.5}{0.5}{0.5}{0.25}{0.25}{0.25}{0.25}
			\NoId{0+\x}{0.5}
			\TenW{W}{1+\x}{0.5}{0.5}{0.5}{0.25}{0.25}{0.25}{0.25}
			\tikzDotsH{2.25+\x}{0.75}
			\TenW{W}{3+\x}{0.5}{0.5}{0.5}{0.25}{0.25}{0.25}{0.25}
			\TenW{-'}{4-0.25+\x}{0.5}{0.5}{0.5}{0}{0}{0}{0}
			\trten{+1+\x}{0.25}{+1+\x}{0.75}
			\trten{+3+\x}{0.25}{+3+\x}{0.75}
			\node[] at (0.25+\x,-0.25) {$j$};
			\node[] at (1.25+\x,-0.25) {$j+1$};
			\node[] at (3.25+\x,-0.25) {$L$};
		\end{tikzpicture}
	\end{center}
	and the following notation for projecting out the identity component in the physical channel \eqref{ksdksnfjsnfjnfjeo} has been used 
	\begin{center}
		\begin{tikzpicture}
			\TenW{W_{ \id}}{8}{0.5}{0.5}{0.5}{0.25}{0.25}{0.25}{0.25}
			\NoId{8}{0.5}
			\node[] at (9,0.715) {=};
			\TenW{W}{9.5}{0.5}{0.5}{0.5}{0.25}{0.25}{0.25}{0.25}
			\node[] at (10.5,0.75) {$-$};
			\TenW{W}{11.5}{0.5}{0.5}{0.5}{0.25}{0.25}{0.25}{0.25}
			\trten{+11.5}{0.25}{+11.5}{0.75}
			\draw[->,thick] (11,0.25)--(11,1.25);
			\node[above] at (12.5,0.45) {.};
		\end{tikzpicture}
	\end{center}
	Now, note that one can rewrite \eqref{ndjsjlsenensjsj} as
	\begin{align}
		\vec{D}_{\mathbf{0},j}=\left(c_{\tilde{\Psi}_j} \tilde{D}^{\smallcancel{\id}}_{0,j} \vec{K}^{-}_0\right.&+\frac{W^{\smallcancel{\id}}_{\mathbf{0},j}}{N}\sum^\SS_{k=j+1} c_{\tilde{\Psi}_k} N^{j+1-k} \left(\tr_{1}(W_{\mathbf{0},1})\right)^{k-(j+1)}\tr_{1}(\tilde{D}_{\mathbf{0},1}) \vec{K}^{-}_\mathbf{0}\notag\\
		&+\left.N^{j-\SS}W^{\smallcancel{\mathbbm{1}}}_{0,j} \left(\tr_{1}(W_{\mathbf{0},1})\right)^{\SS-j}\vec{K}^{(-)'}_\mathbf{0}\right)\notag\\
		=\left( c_{\tilde{\Psi}_j}\tilde{D}^{\smallcancel{\id}}_{0,j} \vec{K}^{-}_0\right.&+\frac{W^{\smallcancel{\id}}_{\mathbf{0},j}}{N}\sum^{\SS-j-1}_{k=0}
		c_{\tilde{\Psi}_{k+j+1}}  N^{-k} \left(\tr_{1}(W_{\mathbf{0},1})\right)^{k}\tr_{1}(\tilde{D}_{\mathbf{0},1}) \vec{K}^{-}_\mathbf{0}\label{ndjsjlse}\\
		&+\left.N^{j-\SS}W^{\smallcancel{\mathbbm{1}}}_{0,j} \left(\tr_{1}(W_{\mathbf{0},1})\right)^{\SS-j}\vec{K}^{(-)'}_\mathbf{0}\right)\notag\,.
	\end{align}
	One can further simplify the evaluation when ones diagonalizes  $\tr(W_{\mathbf{0},1})=S_{\tr(W)}\Lambda_W S^{-1}_{\tr(W)}$
	\begin{align}
		\vec{D}_{\mathbf{0},j}=\left( c_{\tilde{\Psi}_j}\tilde{D}^{\smallcancel{\id}}_{0,j} \vec{K}^{-}_0\right.&+\frac{c_{\tilde{\Psi}_{j+1}} }{N} W^{\smallcancel{\id} }_{\mathbf{0},j} S_{\tr(W)}
		\left(\sum^{\SS-j-1}_{k=0}
		\left(\UniF(\tfrac{p}{2})\UniF(-\tfrac{p}{2})\right)^{-k} N^{-k}
		\Lambda_{W}^{k} \right)S^{-1}_{\tr(W)}\tr_{1}(\tilde{D}_{\mathbf{0},1}) \vec{K}^{-}_\mathbf{0}\notag\\
		&+\left.N^{j-\SS}W^{\smallcancel{\mathbbm{1}}}_{0,j} \left(\tr_{1}(W_{\mathbf{0},1})\right)^{\SS-j}\vec{K}^{(-)'}_\mathbf{0}\right)\notag
	\end{align}
	Unfortunately, one cannot simplify the sum in the last line by the finite geometric series as the residual of the summand is not necessarily invertible for all cases of $R$. However, the above form of $\vec{D}$ is already very efficient for numerical evaluation. 
	After setting the stage, one can now actually compute the Hilbert-Schmidt norms of $\Psi_j$ by evaluating
	\begin{align}\label{mfkdnfnejeme}
		\begin{aligned}
			|\!|\Psi_j|\!|^2=&N^{-j}\, \tr_j\big(\Psi^\dagger_j\, \Psi_j\big)\,,\\
			=&\mathcal{N}^2\cdot N^{-j} \vec{K}^{+,*}_{\mathbf{p}_1} [W^*]^{\mathbf{p}_1,\gamma_1}_{\mathbf{p}_2,\alpha_1}[W^*]^{\mathbf{p}_2,\gamma_2}_{\mathbf{p}_3,\alpha_2}\dots [W^*]^{\mathbf{p}_{j-1},\gamma_1}_{\mathbf{p}_j,\alpha_1} [\vec{D}^*]^{\mathbf{p}_j, \gamma_j}_{\,\,\quad \alpha_j}\\
			&\qquad \quad \times  \vec{K}^{+}_{\mathbf{k}_1} [W]^{\mathbf{k}_1,\gamma_1}_{\mathbf{k}_2,\alpha_1}[W]^{\mathbf{k}_2,\gamma_2}_{\mathbf{k}_3,\alpha_2}\dots [W]^{\mathbf{k}_{j-1},\gamma_1}_{\mathbf{k}_j,\alpha_1} [\vec{D}]^{\mathbf{k}_j, \gamma_j}_{\,\,\quad \alpha_j}
		\end{aligned}
	\end{align}
	or diagrammatically
	\begin{center}
		\begin{tikzpicture}
			\node[above] at (-1.85,0.8) {$|\!|\Psi_j|\!|^2=\mathcal{N}^2\,\cdot$};
			\TenW{+}{-1+0.25}{0.5}{0.5}{0.5}{0}{0}{0}{0}
			\TenW{W}{0}{0.5}{0.5}{0.5}{0.25}{0.25}{0.25}{0.25}
			\TenW{W}{1}{0.5}{0.5}{0.5}{0.25}{0.25}{0.25}{0.25}
			\tikzDotsH{2.25}{0.75}
			\TenW{W}{3}{0.5}{0.5}{0.5}{0.25}{0.25}{0.25}{0.25}
			\TenW{D_j}{4}{0.5}{0.5}{0.5}{0.25}{0.25}{0.25}{0.25}
			\TenW{-}{5-0.25}{0.5}{0.5}{0.5}{0}{0}{0}{0}
			\def\y{0.75}
			
			\TenW{+^*}{-1+0.25}{0.5+\y}{0.5}{0.5}{0}{0}{0}{0}
			\TenW{W^\dagger}{0}{0.5+\y}{0.5}{0.5}{0.25}{0.25}{0.25}{0.25}
			\TenW{W^\dagger}{1}{0.5+\y}{0.5}{0.5}{0.25}{0.25}{0.25}{0.25}
			\tikzDotsH{2.25}{0.75+\y}
			\TenW{W^\dagger}{3}{0.5+\y}{0.5}{0.5}{0.25}{0.25}{0.25}{0.25}
			\TenW{D_j^\dagger}{4}{0.5+\y}{0.5}{0.5}{0.25}{0.25}{0.25}{0.25}
			\TenW{-^*}{5-0.25}{0.5+\y}{0.5}{0.5}{0}{0}{0}{0}
			\trten{0}{0.25}{0}{1.5}
			\trten{1}{0.25}{1}{1.5}
			\trten{3}{0.25}{3}{1.5}
			\trten{4}{0.25}{4}{1.5}
		\end{tikzpicture}
	\end{center}
	where (and analogue for the other appearing tensors)
	\begin{center}
		\begin{tikzpicture}
			
			\TenW{W^\dagger}{-3}{0}{0.5}{0.5}{0.25}{0.25}{0.25}{0.25}
			\node[below] at (0.25,-0.2) {$\beta$};
			\node[right] at (0.75,0.26) {$\mathbf{k}$};
			\node[right] at (-0.7,0.26) {$\mathbf{j}$};
			\node[below] at (0.25,1.2) {$\alpha$};
			\node[] at (-1.25,0.24) {$=$};
			\TenW{W^*}{0}{0}{0.5}{0.5}{0.25}{0.25}{0.25}{0.25}

			\node[below] at (0.25-3,-0.2) {$\alpha$};
			\node[right] at (0.75-3,0.26) {$\mathbf{k}$};
			\node[right] at (-0.7-3,0.26) {$\mathbf{j}$};
			\node[below] at (0.25-3,1.2) {$\beta$};
		\end{tikzpicture}
	\end{center}
Let us introduce the following Dirac notations 
\begin{align}
	 \bra{\mathcal{K}}&= \sum_{\mathbf{k}_1,\mathbf{k}_2}\vec{K}^{+,*}_{\mathbf{k}_1} \vec{K}^{+}_{\mathbf{k}_2}\bra{\mathbf{k}_1,\,\mathbf{k}_2}\,,\\
	 \mathcal{W}&=\sum_{\alpha_j,\,\gamma_j,\,\mathbf{k}_1,\,\mathbf{k}_2}[W^*]^{\mathbf{p}_j,\gamma_j}_{\mathbf{p}_{j+1},\alpha_j}[W]^{\mathbf{k}_j,\gamma_j}_{\mathbf{k}_{j+1},\alpha_j}
	\ket{\mathbf{p}_j,\mathbf{k}_j}\bra{\mathbf{p}_{j+1},\mathbf{k}_{j+1}}\,,\\
	 \ket{\mathcal{D}_j(L)}&= \sum_{\alpha_j,\,\gamma_j,\mathbf{k}_1,\,\mathbf{k}_2}	[\vec{D}^*(L)]^{\mathbf{k}_1, \gamma_j}_{\,\,\quad \alpha_j}[\vec{D}(L)]^{\mathbf{k}_2, \gamma_j}_{\quad\,\,\alpha_j} \ket{\mathbf{k}_1,\,\mathbf{k}_2}\,.
\end{align}
Then one can rewrite \eqref{mfkdnfnejeme} as -- where I have restored for  clarity all dependence on the system size
\begin{align}
	|\!|\Psi_k|\!|^2=\frac{\mathcal{N}^2(L)}{N^k}\,\bra{ \mathcal{K} } \mathcal{W}^{k-1} \ket{\mathcal{D}_k(L)}\,.
\end{align}
This formulation is highly efficient for numerical evaluation, as it reduces the computation to matrix multiplications, where $\mathcal{W}$ is a sparse matrix. \\
\textbf{Localization: }Now, one can employ the normalisation condition $|\!|\mathbbG{\Psi}|\!|^2=1$ to argue for an exponential decay of the HS norm 
\begin{align}
	1=&\lim_{L\to \infty} \sum^L_{k=1}|\!|\Psi_k|\!|^2\\ 
	=&\lim_{L\to \infty} \sum^L_{k=1} \frac{\mathcal{N}^2(L)}{N^k}\,\bra{ \mathcal{K} } \mathcal{W}^{k-1} \ket{\mathcal{D}_k(L)}\\
	=&\sum^{N^4-1}_{\alpha=0} \lambda^{-1}_\alpha \braket{\mathcal{K}\,|\, \mathcal{R}_\alpha } \, \lim_{L\to \infty}\, \sum^L_{k=1}\, \left(\frac{\lambda_\alpha}{N}\right)^k\, \underbrace{\braket{\mathcal{L}_\alpha\,|\, \mathcal{D}_k(L)}\mathcal{N}^2(L)}_{f_L(k,\alpha)}\label{fnsk3mepdmrnw}
\end{align}
where the spectral decomposition $\mathcal{W}=\sum^{N^4-1}_{\alpha=0}\,\lambda_\alpha\, \ket{\mathcal{R}_\alpha}\bra{\mathcal{L}_\alpha}$ with $|\lambda_i|>|\lambda_j|$ for $j>i$ has been used. One sees that the behavior of $\ket{\mathcal{D}_k(L)}$ in respect to $L$ dictates the needed normalization $\mathcal{N}(L)$: for the latter sum in \eqref{fnsk3mepdmrnw} to be convergent and non-vanishing, the limit of the function $f_L(k,\alpha)$ needs to be 
\begin{align}
	\lim_{L\to\infty} f_L(k,\alpha)\sim \mu^k_\alpha\,, \qquad \text{with}\qquad \Big|\frac{\lambda_\alpha\mu_\alpha}{N}\Big|<1
\end{align}
with at least one $\lambda_\alpha\mu_\alpha\neq0$. Now it turns out -- based on numerical investigations for all considered models -- that  
\begin{align}
\Big|\frac{\lambda_0 \mu_0}{N}\Big|=1\, \qquad \text{and} \qquad 0<\Big|\frac{\lambda_\alpha \mu_\alpha}{N}\Big|<1 \qquad \text{with} \qquad \chi>\alpha>0\,, \qquad \chi\ge1	\,.
\end{align}
Therefore, the only way to preserve normalization is to completely get rid-off  the $\alpha=0$ channel by imposing
	\begin{align}\label{jfejfslslslslslslsllejjj}
	\langle \mathcal{K} \,|\, \mathcal{R}_0 \rangle=0\,.
\end{align}
This in turn induces exponential decay of the HS norms:
\begin{align}
 \lim_{L\to \infty}	|\!|\Psi_k|\!|^2=&\lim_{L\to \infty} \frac{\mathcal{N}^2(L)}{N^k}\,\bra{ \mathcal{K} } \mathcal{W}^{k-1} \ket{\mathcal{D}_k(L)}\\
 =&\sum^{N^4-1}_{\alpha=0} \lambda^{-1}_\alpha \braket{\mathcal{K}\,|\, \mathcal{R}_\alpha } \, \lim_{L\to \infty}\, \left(\frac{\lambda_\alpha}{N}\right)^k\, \underbrace{\braket{\mathcal{L}_\alpha\,|\, \mathcal{D}_k(L)}\mathcal{N}^2(L)}_{f_L(k,\alpha)}\\
  =&\sum^{N^4-1}_{\alpha=0} \lambda^{-1}_\alpha \braket{\mathcal{K}\,|\, \mathcal{R}_\alpha } \, \left(\frac{\lambda_\alpha \mu_\alpha}{N}\right)^{k}\\
    =&\sum^{\chi}_{\alpha=1} \lambda^{-1}_\alpha \braket{\mathcal{K}\,|\, \mathcal{R}_\alpha } \, \Big(\underbrace{\frac{\lambda_\alpha \mu_\alpha}{N}}_{|\cdot|<1}\Big)^{k}
\end{align}
Hence, one can refer to \eqref{jfejfslslslslslslsllejjj} as a sufficient condition for localization of the ESZM on the boundary. One finds that $\ket{\mathcal{R}_0}$  is the same for $A^{(1)}_{n-1}$, $A^{(2)}_{2n-1}$, $C^{(1)}_{n}$, $D^{(1)}_{n}$ \emph{provided} $|\Delta|>1$ and is given by the analytical formula
	\begin{align}\label{nfjnfjgnjgn}
		\ket{\mathcal{R}_0}=\frac{1}{N} \left(\sum^{N}_{j=1} \, \ket{j,j}\, \right)\otimes \left(\sum^{N}_{j=1} \, \ket{j,j}\, \right)\,.
	\end{align}
	Having this state explicitly, suitable BCs obeying the orthogonality relation can then be easily identified. Examples are presented in the end matter. However, for the case $B^{(1)}_n$  one has that $\ket{\mathcal{R}_0}$ is given by \eqref{nfjnfjgnjgn}
	for $\Delta>1$ and for $\Delta<\Delta_{B^{(1)}_n}$. Here $\Delta_{B^{(1)}_n}$ are fixed values only depending on $n$, which seem to tend to $-1$ for $n\to\infty$. I do not have an analytical formula for $\Delta_{B^{(1)}_n}$.  I list the values of $\Delta_{B^{(1)}_n}$ in  table \ref{hksmkfnsknfksnf} for the first few $n$.  Also, I have studied the various K-matrices found in \cite{Lima_Santos_All}, but I did not find any $\bra{\mathcal{K}}$ orthogonal to the numerical obtained $\ket{\mathcal{R}_0}$ for $\Delta_{B^{(1)}_n}<\Delta<-1$ as, quite surprisingly, the overlap turned out to be strictly positive independent of the free complex parameters $\{\beta^+\}$. 
	
	\begin{table}[h!]
		\centering
		\begin{tabular}{c c}
			\hline
			$n$ & $\Delta_{B^{(1)}_n}$ \\
			\hline
			2 & -1.80437 \\
			3 &  -1.47576 \\
			4 & -1.3515 \\
			\hline
		\end{tabular}
		\caption{\label{hksmkfnsknfksnf}Numerical obtain values for the bound $\Delta_{B^{(1)}_n}$ for small system sizes.}
	\end{table}

	\begin{figure}
		\begin{center}
			\begin{tikzpicture}
				\def\x{0}
				\def\y{0}
				\node[above] at (\x,\y) {\includegraphics[width=0.45\linewidth]{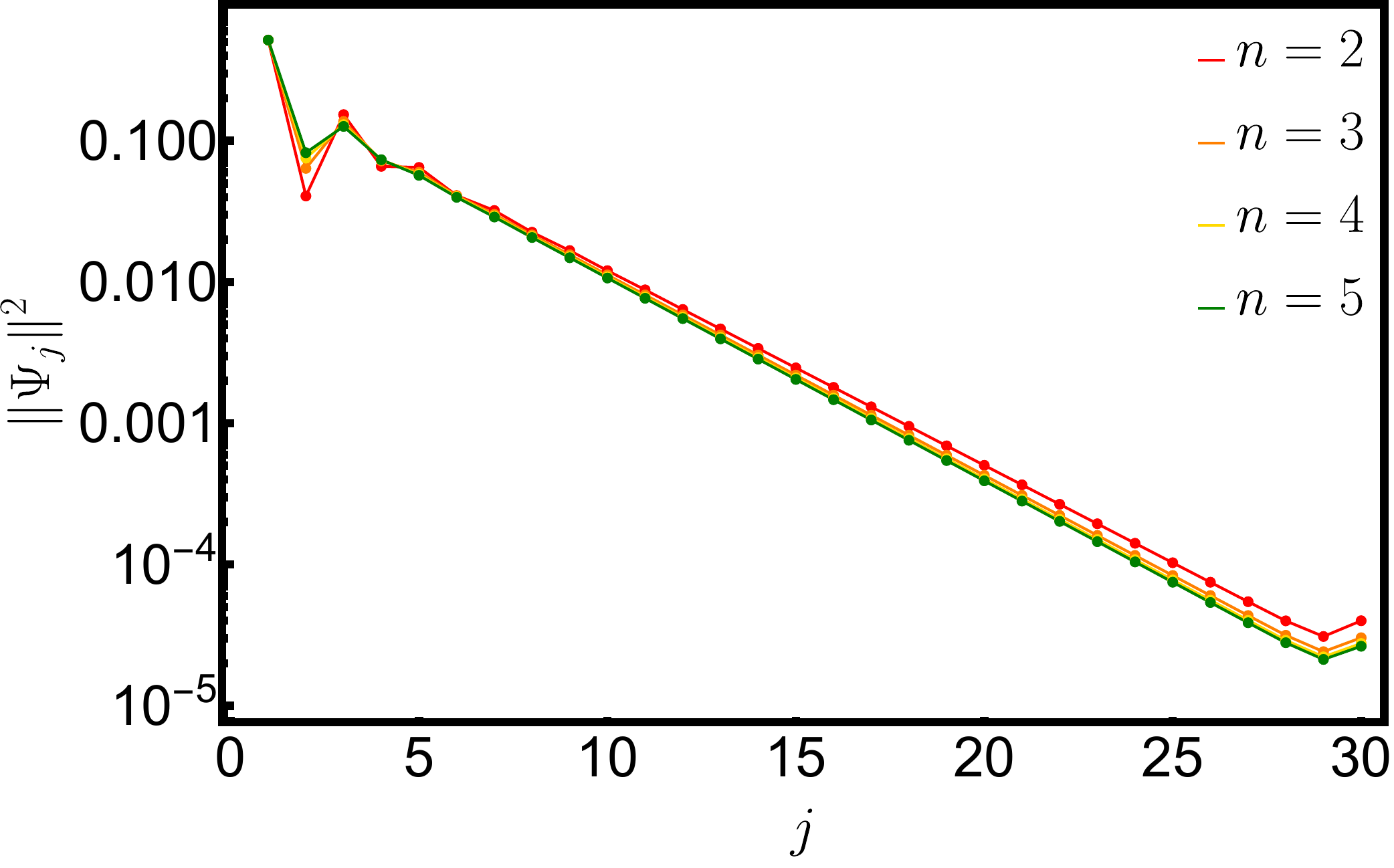}};
				\node[below] at (-3.85+\x,5.35+\y) {$(a)$};
				
				\def\x{8.5}
				\def\y{0}
				\node[above]  at (\x,\y) {\includegraphics[width=0.45\linewidth]{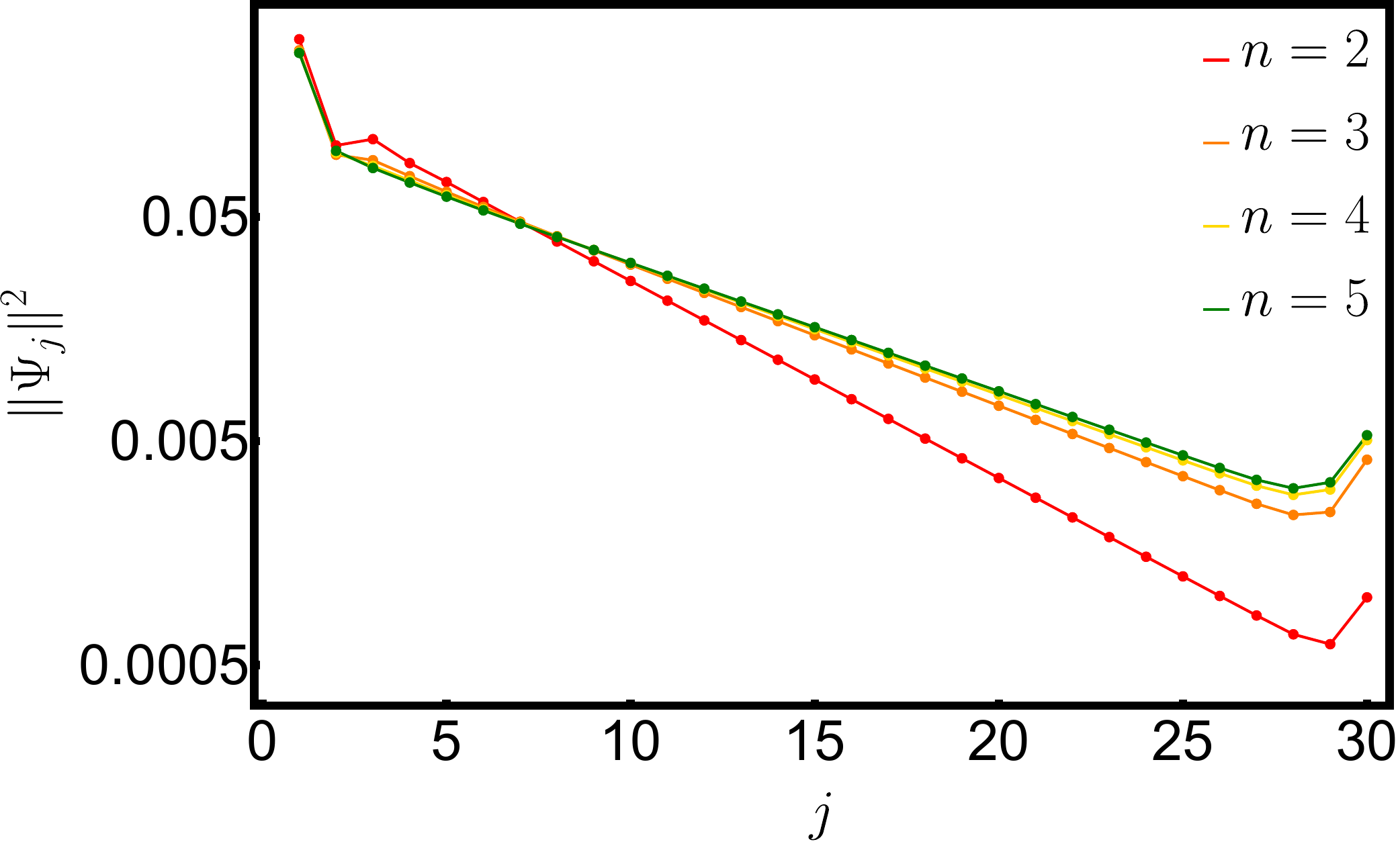}};
				\node[below] at (-3.85+\x,5.35+\y) {$(b)$};
				
				\def\x{0}
				\def\y{-5.25}
				\node[above]  at (\x,\y) {\includegraphics[width=0.45\linewidth]{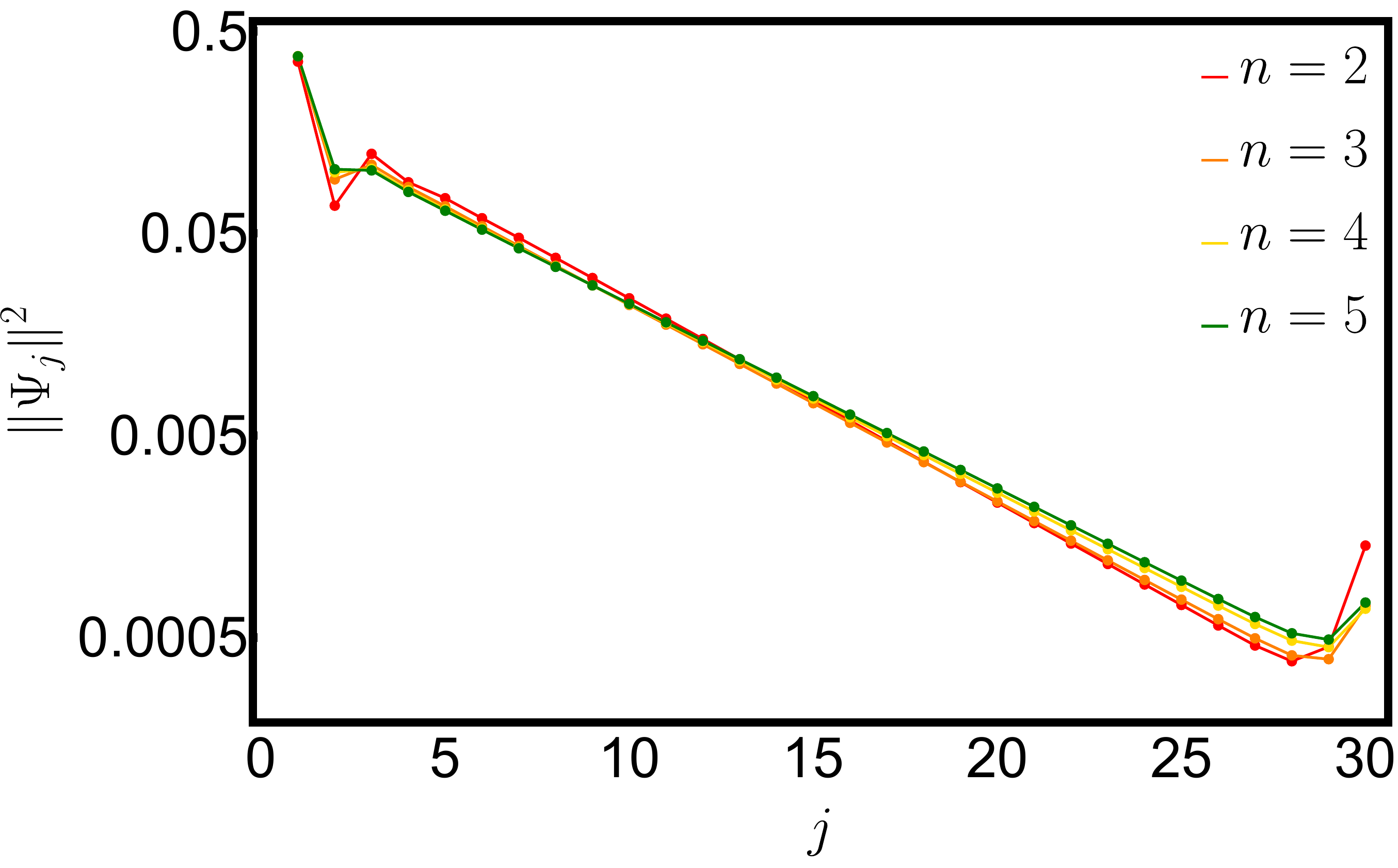}};
				\node[below] at (-3.85+\x,5.35+\y) {$(c)$};
				
				\def\x{8.5}
				\def\y{-5.25}
				\node[above]  at (\x,\y) {\includegraphics[width=0.45\linewidth]{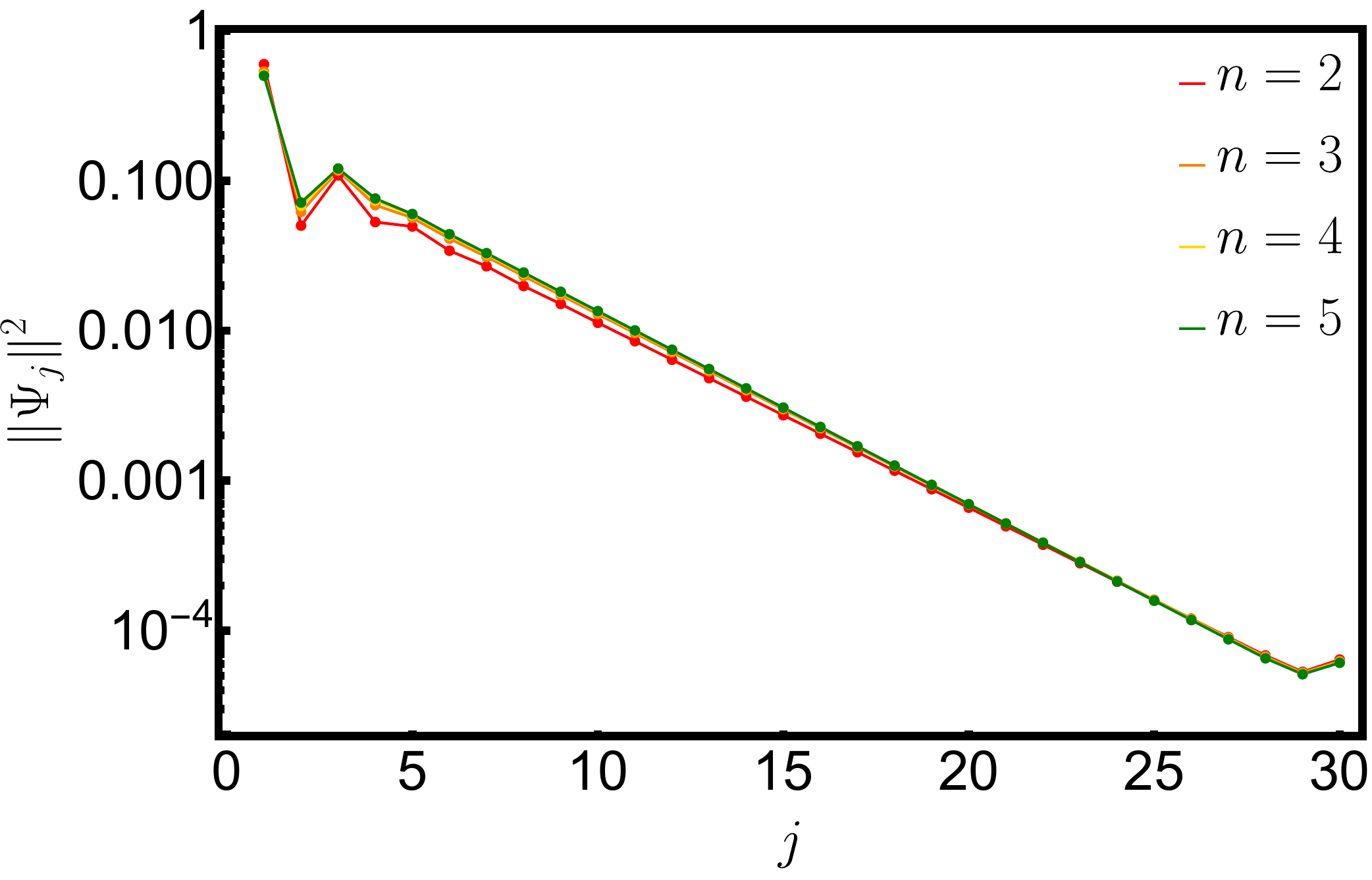}};
				\node[below] at (-3.85+\x,5.35+\y) {$(d)$};
				
				\def\x{4.25}
				\def\y{-10.5}
				\node[above]  at (\x,\y) {\includegraphics[width=0.45\linewidth]{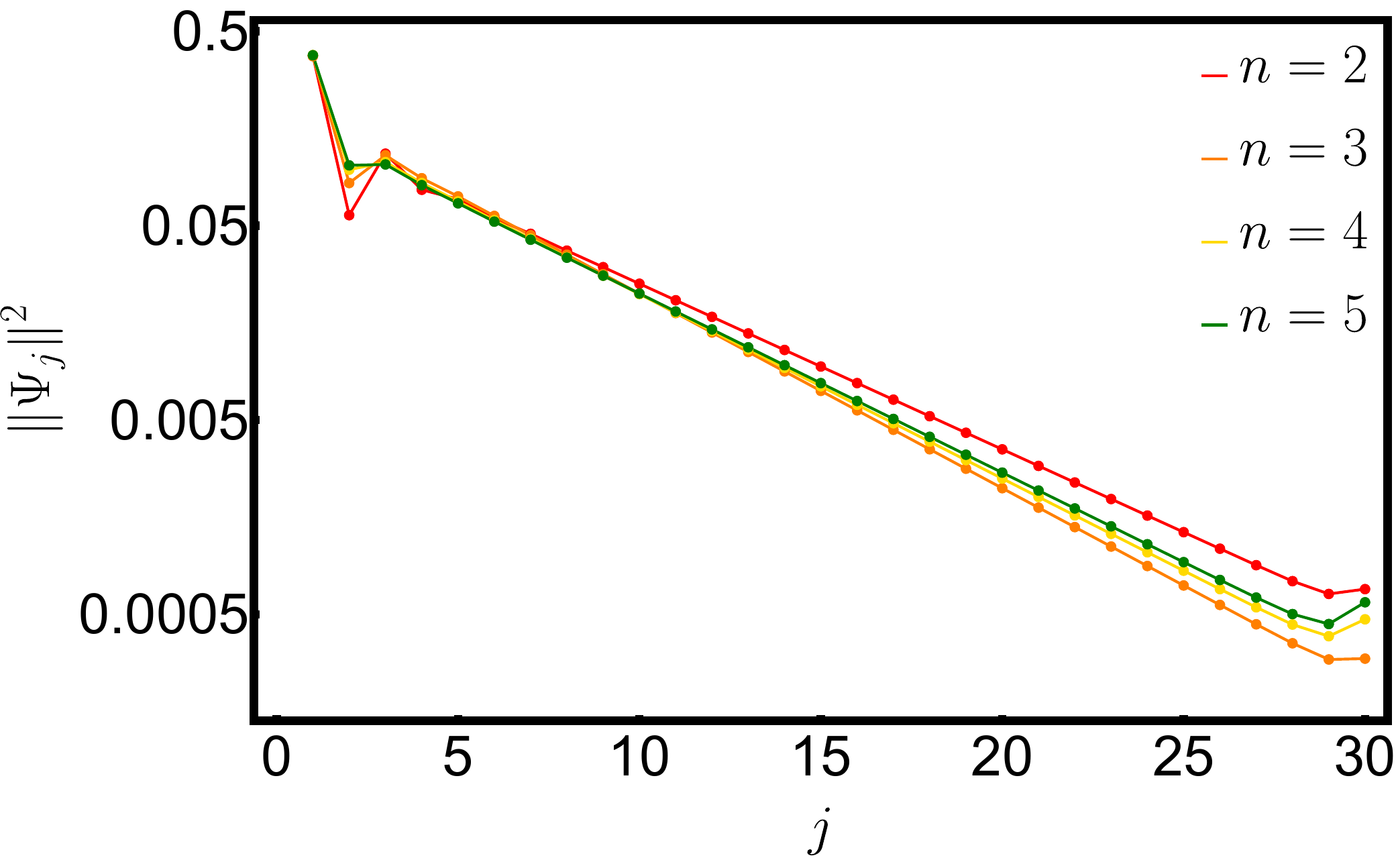}};
				\node[below] at (-3.85+\x,5.35+\y) {$(e)$};
			\end{tikzpicture}
		\end{center}

		\caption{
			Spatial localization of the ESZM of different integrable models for the associated choice of K-matrices.  For all plots,  I fix $K^+$ by isomorphism \eqref{elsnghdlrmgldjshsh} from $K^-$ and $L=30$. One can clearly see an exponential decay. The small spike for sites $j\sim L$ is a small boundary effect. 
			(a)
			$A^{(1)}_{n-1}$,
			$\Delta=1.888$, 
			$K^-$ chosen as \eqref{fndjnfjdnfjdnfjd1},
			$\beta^{+}_{1,1}=(1+q^{2-n})^{-1}$, $\beta^{-}_{1,1}=0$
			(b) $A^{(2)}_{2n-1}$,
			$\Delta=1.433$, 
			$K^-$ chosen as \eqref{eoksnfnfhhs} and
			$\beta^+_{1,1}=-1$, $\beta^-_{1,1}=-0.15$.
			(c) $B^{(1)}_{n}$, 
			$\Delta=1.604$, 
			$K^-$ chosen as \eqref{skdskdnsjjf} 
			and $\beta^+_{1,1}=1$, $\beta^-_{1,1}=0.3$.
			(d) $C^{(1)}_{n}$,
			$\Delta=1.811$ and 
			$K^-$ chosen as \eqref{kejdhfkdoshhd} with $\beta^+_{1,1}=-1$ and $\beta^-_{1,1}=0.15$.
			(e) $D^{(1)}_{n}$,
			$\Delta=1.604$ and 
			$K^-$ chosen as  \eqref{kejdhfkdjdjdjoshhd1} or \eqref{kejdhfkdjdjdjoshhd2} and
			$\beta^+_{1,1}=-1$, $\beta^+_{2,2}=0.2$ and $\beta^-_{1,1}=0.75$, $\beta^-_{2,2}=0.32$.
		}
		\label{fig:Loc_ABCD}
	\end{figure}
\end{document}